\documentclass[superscriptaddress,groupedaddress,nofootnoteinbib,11pt]{article}
\pdfoutput=1

\usepackage{jheppub}

\usepackage{amsmath, amsfonts, amsthm, amssymb, graphicx, color, hyperref}
\usepackage{subcaption}
\usepackage[utf8]{inputenc}

\allowdisplaybreaks[3]

\def \bal#1\eal  {\begin{align} #1 \end{align}}
\def\({\left(}
\def\){\right)}
\def\[{\left[}
\def\]{\right]}
\def\<{\langle}
\def\>{\rangle}
\def\d{\mathrm{d}}

\newcommand{\f}[2]{\frac{#1}{#2}}
\newcommand{\bim} {\begin{itemize}[noitemsep]}

\newcommand{\eim}{\end{itemize}}
\newcommand{\be} {\begin{equation}}
\newcommand{\ee} {\end{equation}}
\newcommand{\bc}{\begin{center}}
\newcommand{\ec}{\end{center}}

\newcommand{\nn} {\nonumber\\}

\newcommand{\mc} {\mathcal}

\newcommand{\ai}{{\alpha}}
\newcommand{\bi}{{\beta}}

\newcommand{\ri}{{\rho}}

\newcommand{\Li}{\Lambda}

\title{On Capped Higgs Positivity Cone}

\author[a]{Dong-Yu Hong,}
\author[a]{Zhuo-Hui Wang}
\author[a,b]{and Shuang-Yong Zhou}

\affiliation[a]{Interdisciplinary Center for Theoretical Study, University of Science and Technology of China, Hefei, Anhui 230026, China}
\affiliation[b]{Peng Huanwu Center for Fundamental Theory, Hefei, Anhui 230026, China}

\emailAdd{principle@mail.ustc.edu.cn}
\emailAdd{wzh33@mail.ustc.edu.cn}
\emailAdd{zhoushy@ustc.edu.cn}

\preprint{{\small USTC-ICTS/PCFT-23-38}}

\date{\today}

\abstract{
The Wilson coefficients of the Standard Model Effective Field Theory are subject to a series of positivity bounds. It has been shown that, while the positivity part of the UV partial wave unitarity leads to the Wilson coefficients living in a convex cone, further including the non-positivity part caps the cone from above. For the Higgs scattering, a capped positivity cone have been obtained using a simplified, linear unitarity conditions and without utilizing the full internal symmetries of the Higgs scattering. Here we further implement the stronger nonlinear unitarity conditions from the UV, which generically gives rise to better bounds. We show that, for the Higgs case in particular, while the nonlinear unitarity conditions per se do not enhance the bounds, the fuller use of the internal symmetries do shrink the capped positivity cone significantly.
}

\begin{document}

\maketitle
\flushbottom

\section{Introduction}

Effective field theories (EFTs) serve as valuable tools for describing low-energy physics without explicit knowledge of the intricate high energy theory. The effectiveness of an EFT depends heavily on precisely determining its Wilson coefficients, which often poses a significant challenge as there can be numerous of them or they might be rather difficult to measure. Recent developments highlight that the general parameter space for Wilson coefficients is mostly inconsistent with the fundamental principles of S-matrix theory such as causality/analyticity and unitarity, except for a small subspace defined by the positivity bounds (see, for example, \cite{Adams:2006sv, deRham:2017avq, deRham:2017zjm, Arkani-Hamed:2020blm, Bellazzini:2020cot, Tolley:2020gtv, Caron-Huot:2020cmc, Chiang:2021ziz, Sinha:2020win, Zhang:2020jyn, Li:2021lpe, Bellazzini:2014waa, Bellazzini:2016xrt, Bern:2021ppb, Alberte:2020jsk, Tokuda:2020mlf, Caron-Huot:2021rmr, Grall:2021xxm, Du:2021byy, Alberte:2021dnj, Bellazzini:2021oaj, Chowdhury:2021ynh, Chiang:2022ltp, Caron-Huot:2022ugt, Caron-Huot:2022jli, Henriksson:2022oeu, Chiang:2022jep, Albert:2022oes, CarrilloGonzalez:2023cbf, Hong:2023zgm, Li:2023qzs, Paulos:2016but, Paulos:2017fhb, He:2018uxa, He:2021eqn, Karateev:2019ymz, Guerrieri:2020bto, Kruczenski:2020ujw, Guerrieri:2021tak, Guerrieri:2021ivu, Albert:2023jtd, Acanfora:2023axz, Miro:2023bon} and \cite{deRham:2022hpx} for a review).

The positivity bounds have been used to constrain the Standard Model Effective Field Theory (SMEFT) \cite{Zhang:2018shp, Bi:2019phv, Bellazzini:2018paj, Remmen:2019cyz, Zhang:2020jyn, Yamashita:2020gtt, Trott:2020ebl, Remmen:2020vts, Remmen:2020uze, Gu:2020thj, Fuks:2020ujk, Gu:2020ldn, Bonnefoy:2020yee, Li:2021lpe, Davighi:2021osh, Chala:2021wpj, Zhang:2021eeo, Ghosh:2022qqq, Remmen:2022orj, Li:2022tcz, Li:2022rag, Li:2022aby, Altmannshofer:2023bfk, Davighi:2023acq, Ellis:2023zim, Chala:2023xjy, Gu:2023emi}. The SMEFT parameterizes generic new physics beyond the Standard Model (SM) based on the SM field content and symmetries, and has been gaining popularity in both theoretical and experimental communities in the absence of new particle discoveries at the LHC. The SMEFT contains numerous Wilson coefficients particularly from dimension-8 \cite{Li:2020gnx, Murphy:2020rsh} and beyond, as the SM is a theory with many field degrees of freedom. 
 
For a theory with multiple degrees of freedom, the positivity bounds significantly reduce the extensive parameter space. For instance, in vector boson scattering (VBS), the elastic positivity bounds can confine the physical dimension-8 Wilson coefficient space to approximately 2\% of the total space \cite{Zhang:2018shp, Bi:2019phv, Vecchi:2007na}. For the 10D dimension-8 VBS subspace involving only the transverse vector bosons, generalized elastic positivity bounds reduce the viable parameter space to about 0.7\% of the total \cite{Yamashita:2020gtt}. Futhermore, for the $s^2$ amplitude coefficients ($s,t,u$ being the standard Mandelstam variables), the optimal positivity bounds can be obtained by a convex geometry approach~\cite{Yamashita:2020gtt, Zhang:2020jyn, Zhang:2021eeo, Bellazzini:2014waa}. When there are sufficient symmetries in the sub-sector we are interested in, one can use a group-theoretical method to compute the positivity cone, and the extremal rays of the cone in this case can be very useful in reverse engineering the UV theory in the event of an observation of non-zero Wilson coefficients \cite{ Zhang:2020jyn, Zhang:2021eeo, Bellazzini:2014waa}. Generically, with fewer symmetries, one can employ a semi-definite programming (SDP) method to compute the the $s^2$ positivity cone \cite{Li:2021lpe}.

The preceding $s^2$ positivity cones are obtained by using only the positivity part of the UV unitarity conditions. Ref \cite{Chen:2023bhu} has initiated the use of the non-positivity parts of the UV unitarity conditions to constrain the SMEFT coefficients, focusing on the scattering involving only the complex Higgs modes. Building upon the methods introduced in \cite{Tolley:2020gtv, Caron-Huot:2020cmc, Chiang:2022jep}, the numerical bounds of Ref \cite{Chen:2023bhu} are obtained by discretizing the UV scales in the fixed-$t$ dispersion relations and using the null constraints and linear programming to extract the constraints on the Wilson coefficients. Specifically, Ref \cite{Chen:2023bhu} derived a set of linear conditions from (nonlinear) partial wave unitarity, which allows the numerical optimizaiton to be easily carried out with some simple {\tt Mathematica} coding.

In this paper, we will revisit the capped positivity bounds on the SMEFT Higgs sector, making use of the nonlinear unitarity conditions on the imaginary part of the UV amplitudes. We will also more carefully take into account all available symmetries of the SMEFT Higgs Lagrangian. With these improvements, significantly better upper bounds are obtained. The paper is organized as follows. In Section 2, we will first derive the scattering amplitudes and dispersion relations from the SMEFT Lagrangian pertaining to the Higgs, and then introduce the null constraints and the nonlinear UV unitarity conditions we will use in this paper. In Section 3, we briefly set up the numerical optimization scheme for computing the two-sided, optimal numerical bounds. In Section 4, we present our results on the capped Higgs positivity cone, and compare with those obtained in Ref \cite{Chen:2023bhu}. We will see that, carefully taking into account the SMEFT Higgs symmetries, the linear unitarity bounds actually give rise to the same positivity bounds as those from the nonlinear unitarity conditions. We conclude in Section 5. In Appendix A, we will show that the linear unitarity conditions of Ref \cite{Chen:2023bhu} can be derived from the nonlinear unitarity conditions. In Appendix B, we will use a bi-scalar theory as an example to demonstrate that generically the nonlinear unitarity conditions are stronger than the linear ones.

\section{Model and setup}

In this section, we derive the amplitudes for Higgs scattering in the SMEFT and the corresponding fixed-$t$ dispersion relations that are used to extract positivity bounds on the dim-8 Wilson coefficients. Then, we proceed to obtain the so-called null constraints by imposing $st$ crossing symmetries on these dispersion relations, and present the nonlinear unitarity conditions we will use in this paper. Combining these ingredients together, we will derive two-sided bounds for the dim-8 Higgs coefficients in the following sections.

\subsection{Amplitudes and dispersion relations}

In the SMEFT, the Higgs retains the same symmetry as in the Standard Model and is a $SU(2)$ doublet. We shall parameterize the Higgs doublet with two complex fields
\be 
H=\f 1 {\sqrt{2}} 
\begin{pmatrix}
\phi_1    \\
\phi_2  
\end{pmatrix} ,
\ee
and will use $\bar{\rm i}$ to denote the antiparticle of particle i. Thanks to the $SU(2)$ internal symmetry, a generic 2-to-2 Higgs scattering amplitude can be parameterized by the invariant tensors of the $SU(2)$ symmetry
$\mc{M}^{\rm i \bar j k  \bar l}(s,t) = \delta_{\rm i\bar j}\delta_{\rm k\bar l} \ai(s,t) + \delta_{\rm i\bar l}\delta_{\rm \bar j k} \bi(s,t)$. Here we choose all the particles to be all-ingoing for the amplitudes.
The rest amplitudes are related to $\mc{M}^{\rm i \bar j k  \bar l}$ by crossing. The $su$ crossing symmetry implies that $\mc{M}^{\rm i \bar j k \bar l}(s,t)= \mc{M}^{\rm i \bar l k \bar j}(u,t)$, which means that we must have $\ai(u,t)=\bi(s,t)$. Thus we can express the Higgs amplitude as
\bal
\label{amp}
\mc{M}^{\rm i \bar j k \bar l}(s,t)  &= \delta_{\rm i\bar j}\delta_{\rm k\bar l} f(s,t) + \delta_{\rm i\bar l}\delta_{\rm \bar j k} f(u,t) .
\eal
Suppose that below the EFT cutoff the theory is weakly coupled so that we can take the tree level approximation. Then, at low energies, we can parameterized $f(s,t)$ as
\bal
f_{\text{EFT}}(s,t)&= a_1s+a_2t+b_1s^2+b_2st+b_3t^2+c_1s^3+c_2s^2t+\cdots ,
\eal
which is the tree level approximation of $f(s,t)$ in the EFT region. 
We will be interested in constraining the Wilson coefficients of the dimension-8 SMEFT operators that will contribute to the 2-to-2 Higgs scattering. There are three of these operators, all of which contain four derivatives and are parameterized as follows
\bal
\mc{L}_{\rm SMEFT} &\supset C_1 \( D_\mu H^\dagger D_\nu H\)\( D^\nu H^\dagger D^\mu H\) + C_2 \( D_\mu H^\dagger D_\nu H\)\( D^\mu H^\dagger D^\nu H\) \nn 
&+ C_3 \( D^\mu H^\dagger D_\mu H\)\( D^\nu H^\dagger D^\nu H\) ,
\eal
where $D^\mu$ is the gauge covariant derivatives. Matching the $C_i$ coefficients with the amplitude coefficients in $f_{\text{EFT}}(s,t)$, we find that
\bal
C_1=b_3\,,~~~C_2=2b_2-b_3\,,~~~C_3=2b_1-b_3 .
\eal

To make use of the null constraints and partial wave unitarity, we need to derive the dispersion relations where the UV amplitudes are expanded with partial waves. To that end, we shall perform the following partial wave expansion 
\bal
\mc{M}^{1 \bar{1} 2 \bar{2}}(s,t)=16\pi\sum_\ell(2\ell+1)P_\ell\(1+\f{2t}{s}\)a_\ell^{s}(s)  , \\
\mc{M}^{1 2 \bar{1} \bar{2}}(s,t)=16\pi\sum_\ell(2\ell+1)P_\ell\(1+\f{2t}{s}\)a_\ell^{t}(s)  , \\
\mc{M}^{1 \bar{2} 2 \bar{1}}(s,t)=16\pi\sum_\ell(2\ell+1)P_\ell\(1+\f{2t}{s}\)a_\ell^{u}(s) ,
\eal
where $P_\ell(x)$ is the Legendre polynomial and we have defined 
\be
\label{astuaijkl}
a_\ell^{s}(s)=a_\ell^{1 \bar{1} 2 \bar{2}}(s),~~~~ a_\ell^{t}(s)=a_\ell^{1 2 \bar{1} \bar{2}}(s),~~~~a_\ell^{u}(s)=a_\ell^{1 \bar{2} 2 \bar{1}}(s) .
\ee
Note that $a_\ell^{s}$, $a_\ell^{t}$ and $a_\ell^{u}$ are related, as they are all expanded from the same function but with different arguments:
\be
\mc{M}^{1 \bar{1} 2 \bar{2}}(s,t)=f(s,t),~~\mc{M}^{1 2 \bar{1} \bar{2}}(s,t)=f(t,s),~~\mc{M}^{1 \bar{2} 2 \bar{1}}(s,t)=f(u,t) .
\ee
These relations will be taken into account by supplying the dispersion relations with null constraints.
The expansions for the other amplitudes can be related to the above three via $a_\ell^{i j k l}(s)=(-1)^\ell a_\ell^{i j l k}(s)$. (We adopt the convention that the indices i, j, k, l refer to particles, indices ${\rm \bar i, \bar j,\bar k, \bar l}$ refer to anti-particles, and indices $i,j,k,l$ refer to particles or anti-particles.)  More explicitly, we have
\bal
\mc{M}^{1 \bar{1} 2 \bar{2}}(s,u) &=\mc{M}^{1 \bar{1} \bar{2} 2}(s,t) =16\pi\sum_\ell(2\ell+1)P_\ell\(1+\f{2t}{s}\)a_\ell^{1 \bar{1} \bar{2} 2}(s)   \nn
    &=16\pi\sum_\ell(2\ell+1)P_\ell\(1+\f{2t}{s}\)(-1)^\ell a_\ell^{s}(s)  , \\
\mc{M}^{1 \bar{1} 2 \bar{2}}(u,s) & =\mc{M}^{1 \bar{1} \bar{2} 2}(u,t)=\mc{M}^{1 2 \bar{2} \bar{1}}(s,t) 
    =16\pi\sum_\ell(2\ell+1)P_\ell\(1+\f{2t}{s}\)a_\ell^{1 2 \bar{2} \bar{1}}(s)   \nn
    &=16\pi\sum_\ell(2\ell+1)P_\ell\(1+\f{2t}{s}\)(-1)^\ell a_\ell^{t}(s)  ,  \\
\mc{M}^{1 \bar{1} 2 \bar{2}}(t,u)  &=\mc{M}^{1 2 \bar{1} \bar{2}}(u,t)  = \mc{M}^{1 \bar{2} \bar{1} 2}(s,t)  
=16\pi\sum_\ell(2\ell+1)P_\ell\(1+\f{2t}{s}\)a_\ell^{1 \bar{2} \bar{1} 2}(s)   \nn
    &=16\pi\sum_\ell(2\ell+1)P_\ell\(1+\f{2t}{s}\)(-1)^\ell a_\ell^{u}(s) .
\eal

With the above ingredients as well as the Froissart-Martin bound \cite{Froissart:1961ux, Martin:1962rt}, a relative simple use of the residue theorem on the complex $s$ plane for fixed $t$, plus some straightforward algebra, allows us to derive the twice subtracted dispersion relations for the amplitudes (see for example \cite{deRham:2017avq}):
\bal
\label{disperRel}
\sum_{\text{EFT poles}}\!\!\!\! \f{\mc{M}^{i j k l}(\mu,t)}{\mu - s} &= z_0(t) + z_1(t) s + 16\sum_\ell\int_{\Li^2}^{\infty}\d\mu (2\ell+1)\bigg( \f{s^2P_\ell(1+{2t}/{\mu})}{\mu^2(\mu-s)}\ri^{i j k l}_\ell(\mu)  \nn 
&  \hspace{100pt} +\f{u^2P_\ell(1+{2t}/{\mu})}{\mu^2(\mu-u)} \ri^{i l k j}_\ell(\mu) \bigg) ,
\eal
where ``EFT poles" denotes the poles of the amplitudes $\mc{M}^{i j k l}$ in the low energy EFT region and $\Li$ is the EFT cutoff and $\ri^{i j k l}_\ell(\mu)=\text{Im}a^{i j k l}_\ell(\mu)$.  $z_n(t)$ are some functions of $t$ that we will not use in this paper, as we are constraining the coefficients in front of the terms $s^nt^m$ with $n\geq 2$. (Of course, with crossing symmetries, coefficients in $z_n(t)$ can often be related to the $s^{n>2}t^m$ coefficients and thus also be bounded.) Now, \eqref{disperRel} is convergent on both sides of the equality, so we can taylor-expand both sides and match the coefficients in front of  $s^n t^m$, which gives a set of sum rules that will be used to derive the positivity bounds. For example, let us consider the dispersion relation of amplitude $\mc{M}^{1 \bar{1} 2 \bar{2}}(s,t)$. Taylor-expanding both sides of the dispersion relation, we get
\bal
& a_1s + a_2t+ +b_1s^2 + b_2 s t + b_3 t^2 + c_1 s^3 + c_2 s^2 t  + \cdots
 = z_0(t) + \left\langle\f{(\ri^s_\ell(\mu) + \ri^u_\ell(\mu))}{\mu^3}\right\rangle t^2 
 \nonumber\\
&~~~~~~~ + \(z_1(t) + 2\left \langle\f{\ri^u_\ell(\mu)}{\mu^3}\right \rangle t \)s
+ \left \langle\f{(\ri^s_\ell(\mu) + \ri^u_\ell(\mu))}{\mu^3}\right \rangle s^2 
 + \left \langle\f{\ri^s_\ell(\mu) - \ri^u_\ell(\mu)}{\mu^4}\right \rangle s^3 
 \nonumber\\
&~~~~~~~ +  \left \langle\f{(\ell(1 + \ell)\ri^s_\ell(\mu)+ (-3 + \ell + \ell^2)\ri^u_\ell(\mu)}{\mu^4}\right \rangle s^2 t + \cdots  
\eal
where we have defined 
\be 
\<\cdots\>=16\sum_\ell\int_{\Li^2}^{\infty}\d\mu(2\ell+1)\cdots
\ee
and matching the coefficients in front of the terms $s^n t^m$, we can get
\bal
\label{disex3}
b_1 &= \left \langle\f{\ri^s_\ell(\mu) + \ri^u_\ell(\mu)}{\mu^3}\right \rangle ,
\\
\label{nullex1}
c_1 &= \left \langle\f{\ri^s_\ell(\mu) - \ri^u_\ell(\mu)}{\mu^4}\right \rangle   ,
\\
\label{nullex2}
c_2 &= 
\left \langle\f{(\ell(1 + \ell)\ri^s_\ell(\mu)+ (-3 + \ell + \ell^2)\ri^u_\ell(\mu)}{\mu^4}\right \rangle ,
\\
&~~\vdots  \nonumber 
\eal
These sum rules connect the unknown UV amplitudes with the low energy Wilson coefficients. The positivity bounds are the imprints of the UV information on the IR physics, passed down by these dispersion relations/sum rules. 

\subsection{Null constraints}

The fixed $t$ dispersion relations above or the sum rules extracted from them only include part of the full crossing symmetries of the amplitudes. To utilize the full crossing symmetries, we can simply impose the unrealized crossing symmetries, as extra conditions, on these fixed $t$ dispersion relations or the sum rules. This gives rise to null constraints, which can significantly strengthen the positivity bounds, capable to bound the Wilson coefficients from the below and from the above \cite{Tolley:2020gtv, Caron-Huot:2020cmc}.

In the Higgs case, the null constraints can be obtained by equating different expressions of the same Wilson coefficient in various dispersion relations. For example, the coefficients in front of the terms $s^3$ and $s^2t$ in the dispersion relation of $\mc{M}^{1 \bar{1} 2 \bar{2}}(s,t)$ give rise to sum rules for $c_1$ and $c_2$, as shown in (\ref{nullex1}) and (\ref{nullex2}). On the other hand, from the dispersion relation of $\mc{M}^{1 \bar{1} 2  \bar{2}}(t,s)$, the sum rule obtained from the $s^2t$ term is given by
\bal 
\label{nullex3}
-3c_1+2c_2 = \left \langle\f{(1 + (-1)^\ell) \ell (1 + \ell) \ri^t_\ell(\mu) + (-1)^
  \ell (-3 + \ell + \ell^2) (\ri^s_\ell(\mu) + \ri^u_\ell(\mu))}{\mu^4}\right \rangle .
\eal
Plugging the sum rules (\ref{nullex1}) and (\ref{nullex2}) into the sum rule (\ref{nullex3}), we can get one null constraint:
\bal 
\label{nullex4}
0=\left \langle\f{\[3 - 3(-1)^\ell + (-2 + (-1)^\ell)\ell + (-2 + (-1)^\ell)\ell^2\]
  \(\ri^s_\ell+  \ri^u_\ell\)  + (1 + (-1)^\ell)\ell(1 + \ell)  \ri^t_\ell}{\mu^4} \right \rangle .
\eal
To get independent null constraints, we only need to extract sum rules from the dispersion relations of $\mc{M}^{1 \bar{1} 2  \bar{2}}(s,t),\mc{M}^{1 \bar{1} 2  \bar{2}}(t,s)$ and $\mc{M}^{1 1 \bar{1} \bar{1}}(s,t)$. If a coefficient appears in multiple sum rules, we can obtain null constraints as illustrated above. As the order of the sum rules increases, the number of independent null constraints increases, but all of these can be easily handled by a symbolic algebra system.

\subsection{Nonlinear unitarity}
Ref \cite{Chen:2023bhu} derived a set of linearized unitarity conditions that can be used to obtain two-sided bounds on generic dim-8 Wilson coefficients. These linearized unitarity conditions are explicit, simple and easy to use in a linear program. In fact, they can be easily implemented with simple {\tt Mathematica} coding to compute the numerical bounds. In this paper, we further use stronger, nonlinear unitarity conditions, which generally lead to stronger bounds; see Appendix \ref{sec:biS}. For the Higgs case, however, due to the high degrees of the internal symmetries, the nonlinear unitarity conditions are actually equivalent to the linear ones, as we shall see in Section \ref{sec:bound}.

Recall that the full unitarity condition is $\mathbb{S} \mathbb{S}^\dagger=\mathbb{I}$, where $\mathbb{S}$ is the S-matrix and $\mathbb{I}$ is the corresponding identity matrix. If we restrict to a subspace of the space of all outgoing states, the reduced unitarity conditions can be written as $\hat{\mathbb{S}} \hat{\mathbb{S}}^\dagger \preceq  \mathbb{I}$, where $\hat{\mathbb{S}}$ is the projection of ${\mathbb{S}}$ to the subspace. Splitting the projected S-matrix into an identity matrix plus a transfer matrix $\hat{\mathbb{T}}$: $\hat{\mathbb{S}} = \mathbb{I} + i \hat{\mathbb{T}}$, we have $(\mathbb{I}-{\rm Im} \mathbb{T})^2 + ({\rm Re}\mathbb{T})^2\preceq  \mathbb{I}$. Since $({\rm Re}\mathbb{T})^2$ is semi-positive, a weaker but simpler condition is $ \mathbb{I}-(\mathbb{I}-{\rm Im} \mathbb{T})^2 \succeq 0$, which is equal to the following linear matrix inequalities
\be 
\label{nonlinearU}
\text{Im}\mathbb{T} \succeq 0 \,,   \qquad  2 \mathbb{I} - \text{Im}\mathbb{T} \succeq 0 .
\ee
In the scattering, angular momenta are conserved, so the above inequalities also apply to each partial waves:
\be 
\label{nonlinearUell}
\text{Im}\mathbb{T}_\ell \succeq 0 \,,   \qquad  2 \mathbb{I} - \text{Im}\mathbb{T}_\ell \succeq 0 .
\ee
Note that in terms of partial wave amplitudes, these unitarity conditions are highly nonlinear. For the Higgs case we have in hand, for each partial wave, $\mathbb{T}_\ell$ is a $16\times16$ matrix and the partial wave amplitudes are related to it by
\bal
\ri^{ijkl}_{\ell}(s)=
\begin{cases}
\mathbb{T}^{ijkl}_{\ell}(s) /2  &\qquad\text{for $i\ne j$ and $k\ne l$}    \\
\mathbb{T}^{ijkl}_{\ell}(s)   &\qquad\text{for $i=j$ and $k=l$} \\
\mathbb{T}^{ijkl}_{\ell}(s) / \sqrt{2} &\qquad\text{for ($i\ne j$ and $k=l$) or ($i=j$ and $k\ne l$)} .
\end{cases}
\eal 
where the factor 2 comes from the bose symmetry. 

These conditions are generically stronger than those linear unitarity conditions obtained in Ref \cite{Chen:2023bhu}, as demonstrated in Appendix \ref{sec:biS} for the case of a simple bi-scalar theory.  In Appendix \ref{sec:linear}, we show how to re-derive the linear conditions of Ref \cite{Chen:2023bhu} from the nonlinear conditions (\ref{nonlinearUell}).

\section{Numerical implementation}

In the last section, we have derived the sum rules which express the Wilson coefficients in terms of a sum of different UV spin contributions and each UV spin contribution is expressed as an integral over the UV energy scale. We do not know the exact values of the UV partial wave amplitudes, but they should satisfy the partial wave unitarity. Now, we shall numerically implement the nonlinear unitarity conditions \eqref{nonlinearUell} within SDPB. Additionally, the UV partial wave amplitudes should also satisfy the null constraints, which are also easy to implement with the SDPB package\cite{Landry:2019qug}. In this section, we shall set up the numerical method to compute the optimal bounds on the Wilson coefficients.

Our strategy is to discretize the UV scale $\mu$, after which we are left with a finite number of the UV partial wave amplitudes $a^{s}_{\ell}(\mu)$, $a^{t}_{\ell}(\mu)$ and $a^{u}_{\ell}(\mu)$, the decision variables in the optimization problem. Specifically, in the numerical scheme, we choose $\ell=0,1,\dots,\ell_{{M}},\ell_{\infty}$ and $\Li^2/\mu=1/N,\dots,1$, where $\ell_M$ and $N$ are two sufficiently large integers for the numerics to converge and a larger partial wave $\ell_\infty\gg \ell_M$ is chosen to make the numerics converge faster.
For example, under the discretization, \eqref{disex3} becomes
\bal
b_1&= \sum_\ell16(2\ell+1)\int_{\Li^2}^{\infty}\d\mu\f{\ri^s_\ell(\mu) + \ri^u_\ell(\mu)}{\mu^3}  \nn
&\approx \f 1 {\Li^4}\sum_{\ell=0}^{\ell_{\text{M}},\ell_{\infty}}16(2\ell+1)\sum_{n=1}^{N}\f 1 N \f n N\(\ri^s_{\ell,n} + \ri^u_{\ell,n}\) ,
\eal
where we have defined $\ri^s_{\ell,n}=\ri^s_\ell(\Lambda^2 N/n)$ and so on. Note that now $b_1$ is a finite, linear combination of the decision variables. The same discretization is also applied to the null constraints. The null constraints are equality constraints, and in SDPB these equality constraints can be implemented by both imposing $(...)\geq 0$ and $(...)\leq 0$. 
With these setups, we can now propose our semi-definite program to get the bounds on the Wilson coefficients:
\bal
&\text{\textbf{Decision variables}}   \nn
&\qquad \ri^{s}_{\ell,n},~\ri^{t}_{\ell,n},~\ri^{u}_{\ell,n} \text{\qquad for $\ell=0,1,\dots,\ell_{\text{M}},\ell_{\infty}$ and $n= 1,2,..., N$}   \\
&\text{\textbf{Maximize/Minimize}}   \nn
&\qquad\sum_{I=1}^3 \ai_I C_{I},\text{\quad where} \nn
&\qquad C_1=\f 1 {\Li^4}\sum_{\ell=0}^{\ell_{\text{M}},\ell_{\infty}} 16(2 \ell+1)\sum_{n=1}^{N}\f 1 N \f n N \(1-\((-1)^{\ell}\) \(\ri^{s}_{\ell,n} +\ri^{t}_{\ell,n}\)+\((-1)^\ell+1\)\ri^{u}_{\ell,n}\)  \\
&\qquad C_2=\f 1 {\Li^4}\sum_{\ell=0}^{\ell_{\text{M}},\ell_{\infty}} 16(2 \ell+1)\sum_{n=1}^{N}\f 1 N \f n N \(\((-1)^{\ell}-1\) \(\ri^{s}_{\ell,n} +\ri^{u}_{\ell,n}\)+\((-1)^\ell+1\)\ri^{t}_{\ell,n}\)  \\
&\qquad C_3=\f 1 {\Li^4}\sum_{\ell=0}^{\ell_{\text{M}},\ell_{\infty}} 16(2 \ell+1)\sum_{n=1}^{N}\f 1 N \f n N \(\((-1)^{\ell}+1\) \ri^{s}_{\ell,n} +\((-1)^\ell-1\)\(\ri^{t}_{\ell,n}-\ri^{u}_{\ell,n}\)\)  \\
&\text{\textbf{Subject to}}    \nn
&\qquad \text{Unitarity conditions }\eqref{nonlinearU}  \nn
&\qquad \text{Null constraints such as }\eqref{nullex4}   \nonumber
\eal
$\ai_I$ are constants to be chosen by the user, which specifies the direction in the Wilson coefficient space $\{C_I\}$ that one wants to bound. In practice, since some of the unitarity conditions contain constants, we can introduce an extra decision variable and use the SDPB normalization to set this variable to 1.

In this paper, we shall only present 1D and 2D bounds. For the 1D bounds, we calculate the bounds on each of $C_{I}$. To get 2D bounds, we set one of $\ai_I$ to zero and use the angular optimization method to compute the boundary of the bounds. For example, to obtain the bounds on $C_1$ and $C_2$, we set $(\ai_1,\ai_2,\ai_3)=(\cos\theta,\sin\theta,0)$. For each fixed $\theta$, we use the SDPB package to obtain a lower and an upper bound on the objective $\cos\theta C_1 + \sin\theta C_2$, each upper or lower bound delineating a half plane in the $C_1$-$C_2$ space. Doing this for a number of $\theta$, the many half-spaces carve out a 2D boundary in the $C_1$-$C_2$ space.

\section{Bounds on dim-8 Higgs operators}
\label{sec:bound}

In this section, we present the numerical results on the SMEFT Higgs coefficients $C_{1}$, $C_2$ and $C_3$. With the SDP setup in the last section, we can find both upper and lower positivity bounds on them. We will compare our results with those from \cite{Chen:2023bhu}. The positivity bounds obtained in Ref \cite{Chen:2023bhu} are already often stronger than the experiments bounds and the so-called partial wave unitarity bounds. As we see below, our bounds here are even stronger. Note that the partial wave unitarity bounds are {\it not} positivity bounds. The partial wave unitarity bounds reply only on partial wave unitarity within the low energy EFT, and  dispersion relations are not used in their derivation. In comparison, the positivity bounds\,\footnote{We emphasize that, in this paper, we have broaden the definition of positivity bounds, in that we also refer to the bounds obtained by using the non-positivity part of the unitarity conditions as positivity bounds.} (sometimes also known as causality bounds) are built up on the dispersion relations, whose existence relies on causality of the S-matrix, and make use of the partial wave unitarity of the unknown UV theory.

\begin{table}
    \centering
    \begin{tabular}{|c|c|c|c|c|c|c|}
    \hline
                &  \multicolumn{2}{|c|}{$\bar{C_1}=C_1 \Li^4 /(4\pi)^2$}  &  \multicolumn{2}{|c|}{$\bar{C_2}=C_2 \Li^4 /(4\pi)^2$}  &  \multicolumn{2}{|c|}{$\bar{C_3}=C_3 \Li^4 /(4\pi)^2$} \\
    \cline{2-7}
                & lower &  upper & lower & upper& lower & upper\\
    \hline
    Linear      &  $-0.130$&$0.774$  &0  & $0.638$ & $-0.508$ & $0.408$\\
    \hline
    Linear2   &  $-0.086$   &$0.467$   &$0$  &$0.378$  &$-0.387$  &$0.167$ \\
    \hline
    Nonlinear   &  $-0.086$   &$0.467$  &$0$  &$0.378$  &$-0.387$  &$0.167$ \\
    \hline
    \end{tabular}
    \caption{Comparison of positivity bounds on individual coefficients $C_I$ from the linear and nonlinear unitarity conditions. Here, the ``Linear2" and ``Nonlinear" results are our results in this paper obtained using linear and nonlinear unitarity conditions respectively, while the ``Linear" results are from \cite{Chen:2023bhu} using linear unitarity conditions but without using full Higgs symmetries. In this table, the numerical parameters are $N=10$ and $\ell_M=20$, and 42 null constraints are used.}
    \label{tab:1d}
\end{table}

\begin{figure}[tbp]
    \begin{subfigure}[b]{0.3\textwidth}
        \includegraphics[width=\textwidth]{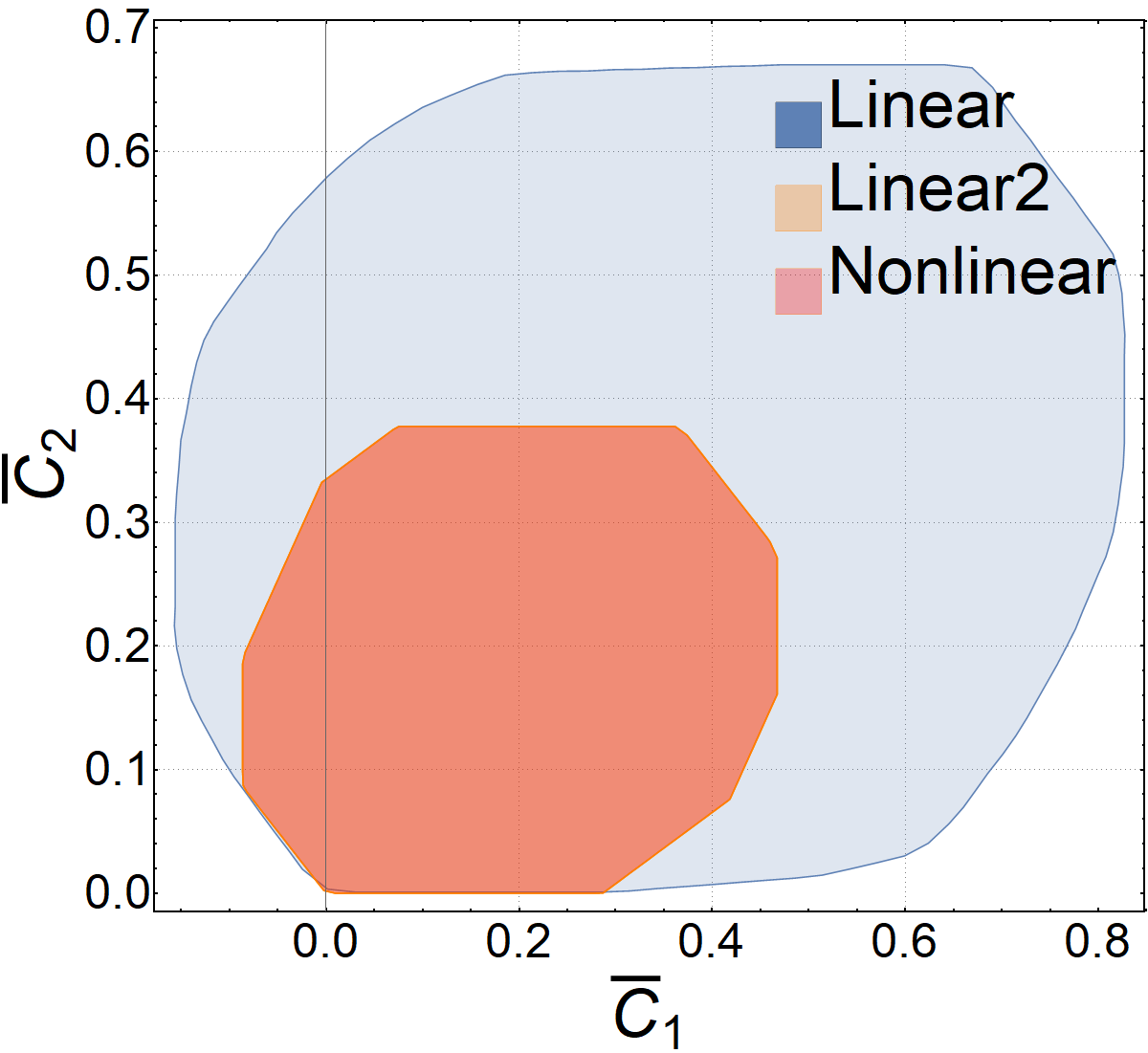}
    \end{subfigure}
    \begin{subfigure}[b]{0.3\textwidth}
        \includegraphics[width=\textwidth]{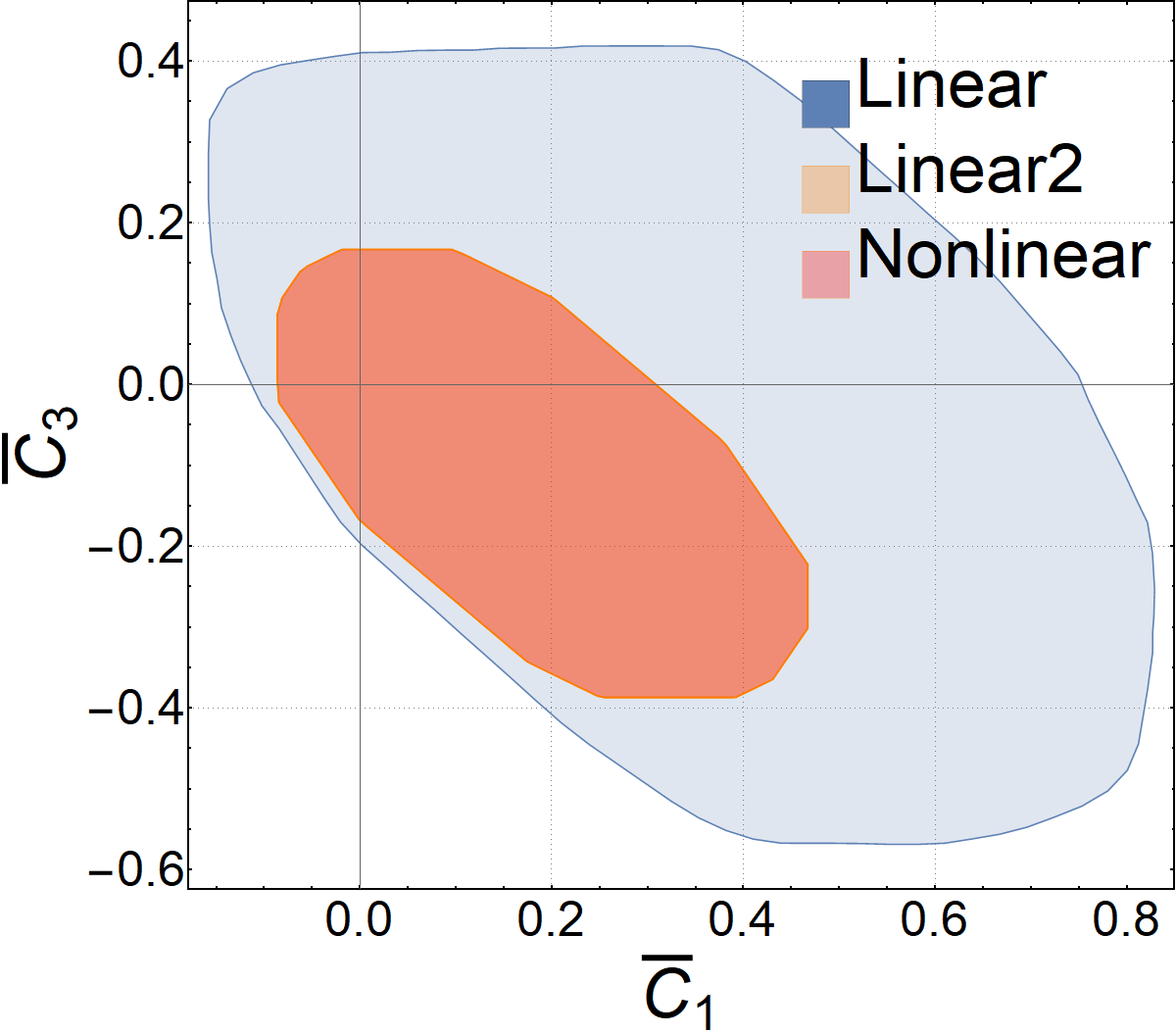}
    \end{subfigure}
    \begin{subfigure}[b]{0.3\textwidth}
        \includegraphics[width=\textwidth]{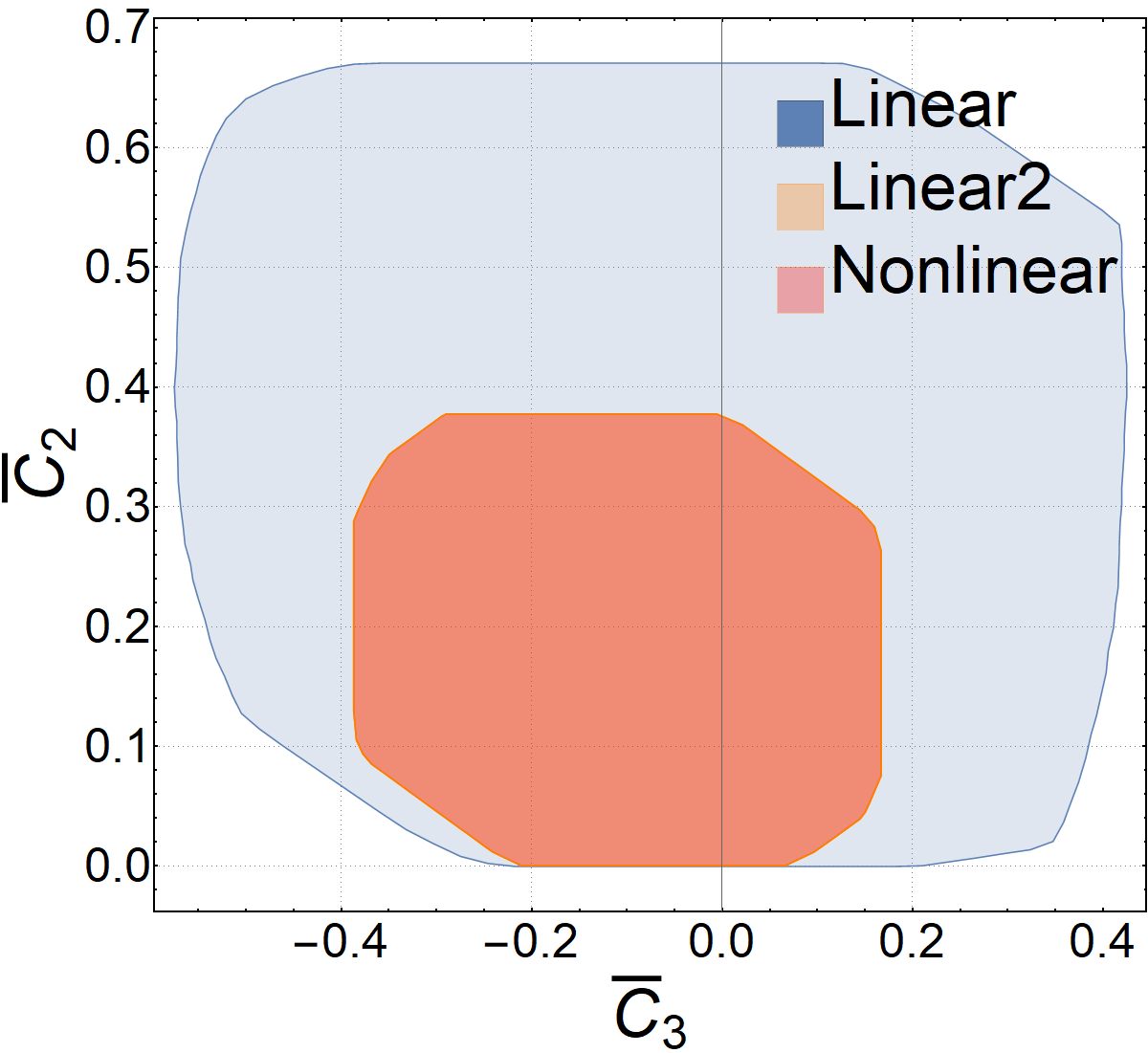}
    \end{subfigure}
    \caption{Positivity regions in the 2D subspaces of $C_1, C_2$ and $C_3$ by using linear and nonlinear unitarity conditions. Here, $\bar{C_i}=C_i \Li^4 /(4\pi)^2$. The orange and red regions are the results of the current paper using linear and nonlinear unitarity conditions respectively, while the blue region are from \cite{Chen:2023bhu}, which uses linear unitarity conditions but without using full Higgs symmetries. The orange and red regions are the same. We choose $N=10,\ell_M=20$ and use 42 null constraints.}
    \label{fig:2d}
\end{figure}

In Table \ref{tab:1d}, we see that, comparing with the results in \cite{Chen:2023bhu}, labeled as ``Linear'', our 1D ``Nonlinear'' bounds are much stronger, almost by a factor of 2. Comparing to the experiments bounds and the partial wave unitarity bounds that have also been obtained in \cite{Chen:2023bhu}, which we shall not repeat here, these new results will be more useful in helping the phenomenological analysis of the collider data. We also compute the 2D positivity bounds in the $C_1$-$C_2$, $C_1$-$C_3$ and $C_2$-$C_3$ plane respectively, which are shown in Figure \ref{fig:2d}. From these plots, we consistently see an improvement by about a factor of 3 or 4, compared to the results of \cite{Chen:2023bhu}. Clearly, in the total 3D parameter space spanned by $C_1$, $C_2$ and $C_3$, the improvement factor is even greater.

%Note that, in Figure \ref{fig:2d}, we are plotting 3 two-dimensional projections of the {\it capped} positivity cone. This is different from the positivity cone of \cite{Zhang:2020jyn} in that the capped positivity cone now additionally has upper bounds, much like an ice cream cone; see Figure 1 of Ref \cite{Chen:2023bhu} for a cartoon explanation. That is, the rays of the cone of \cite{Zhang:2020jyn} go to infinity, while the current capped cone is a set of finite-length line segments. In the two-dimensional projections of the capped positivity cone, we can still see some parts of the \cite{Zhang:2020jyn} cone. For example, in the left plot of Figure \ref{fig:2d}, the projection of the \cite{Zhang:2020jyn} cone is given by the lines of $C_1\ge0$ and $C_1+C_2\ge0$, which pass through the origin, and we see that this cone is now capped from the above by a smooth curve. The reason that we can now bound the cone from above is because: 1) we now have added the null constraints, which uses the information away from the forward limit; 2) we have made fuller use of the partial wave unitarity conditions in the UV. For more details about this, readers are referred to Ref \cite{Chen:2023bhu}.

In Table \ref{tab:1d} and Figure \ref{fig:2d}, the ``Linear2'' bounds are the positivity bounds that can be obtained with the linear unitarity conditions of \cite{Chen:2023bhu} but with all the symmetries of the SMEFT Higgs included. As it happens, these ``Linear2'' positivity bounds are numerically the same as our ``Nonlinear'' bounds. This is coincidental for the case of the SMEFT Higgs, due to the presence of strong internal symmetries. To see why this happens, let us compute the eigenvalues of the matrices ${\rm Im}\mathbb{T}_\ell$ and $2\mathbb{I}-{\rm Im}\mathbb{T}_\ell$, which are given respectively
\bal
&\text{Distinct eigenvalues of ${\rm Im}\mathbb{T}_\ell$ :}   \nn
&\qquad -2(-1 + (-1)^\ell)\ri^t_\ell\,,~~ (1 + (-1)^\ell)\ri^t_\ell\,,~~ 
 2(1 + (-1)^\ell)\ri^t_\ell\,,~~ 2(-1 + (-1)^\ell)\ri^u_\ell\,,~~ \nn &\qquad
 2(1 + (-1)^\ell)\ri^u_\ell\,,~~ 
 2(-1 + (-1)^\ell)(2\ri^s_\ell + \ri^u_\ell)\,,~~ 
 2(1 + (-1)^\ell)(2\ri^s_\ell + \ri^u_\ell) , \\
&\text{Distinct eigenvalues of $2\mathbb{I}-{\rm Im}\mathbb{T}_\ell$}:   \nn
&\qquad2(1 - \ri^t_\ell + (-1)^\ell \ri^t_\ell)\,,~~ -2(-1 + 
    \ri^t_\ell + (-1)^\ell \ri^t_\ell)\,,~~ 
 2 - \ri^t_\ell - (-1)^\ell
   \ri^t_\ell\,,~~ \nn &\qquad -2(-1 - \ri^u_\ell + (-1)^\ell \ri^u_\ell)\,,~~  -2(-1 - 
    2\ri^s_\ell + 2(-1)^\ell \ri^s_\ell - 
    \ri^u_\ell + (-1)^\ell \ri^u_\ell)\,,~~ \nn &\qquad
   -2(-1 + \ri^u_\ell + (-1)^\ell \ri^u_\ell)\,,~~ -2(-1 + 
    2\ri^s_\ell + 2(-1)^\ell \ri^s_\ell + 
    \ri^u_\ell + (-1)^\ell \ri^u_\ell) .
\eal
The semi-positive definiteness of ${\rm Im}\mathbb{T}_\ell$ and $2\mathbb{I}-{\rm Im}\mathbb{T}_\ell$ are just the semi-positivity of these eigenvalues. Remembering \eqref{astuaijkl} and the relation $\ri_\ell^{i j k l}(s)=(-1)^\ell \ri_\ell^{i j l k}(s)$, it is easy to see that the positivity of these eigenvalues exactly give rise to the linear unitarity conditions for the SMEFT Higgs.

Nevertheless, the nonlinear unitarity conditions are in general stronger than the linear unitarity conditions derived in Appendix \ref{sec:linear}. In Appendix \ref{sec:biS}, as a simple example, we show that in a generic $\mathbb{Z}_2$ bi-scalar theory, the nonlinear bounds are indeed stronger than the linear bounds.

\begin{figure}[tbp]
    \begin{subfigure}[b]{0.3\textwidth}
        \includegraphics[width=\textwidth]{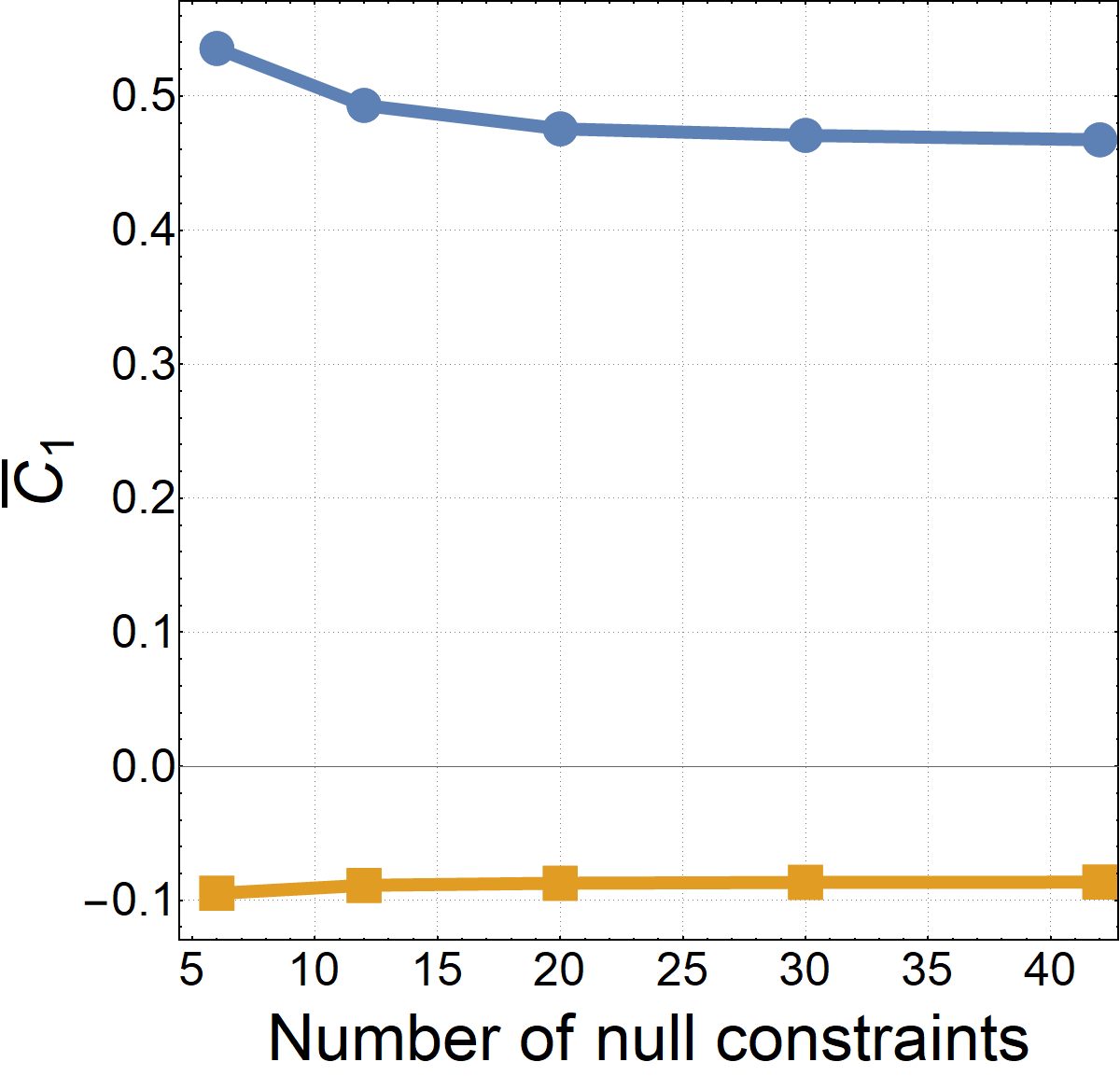}
    \end{subfigure}
    \begin{subfigure}[b]{0.3\textwidth}
        \includegraphics[width=\textwidth]{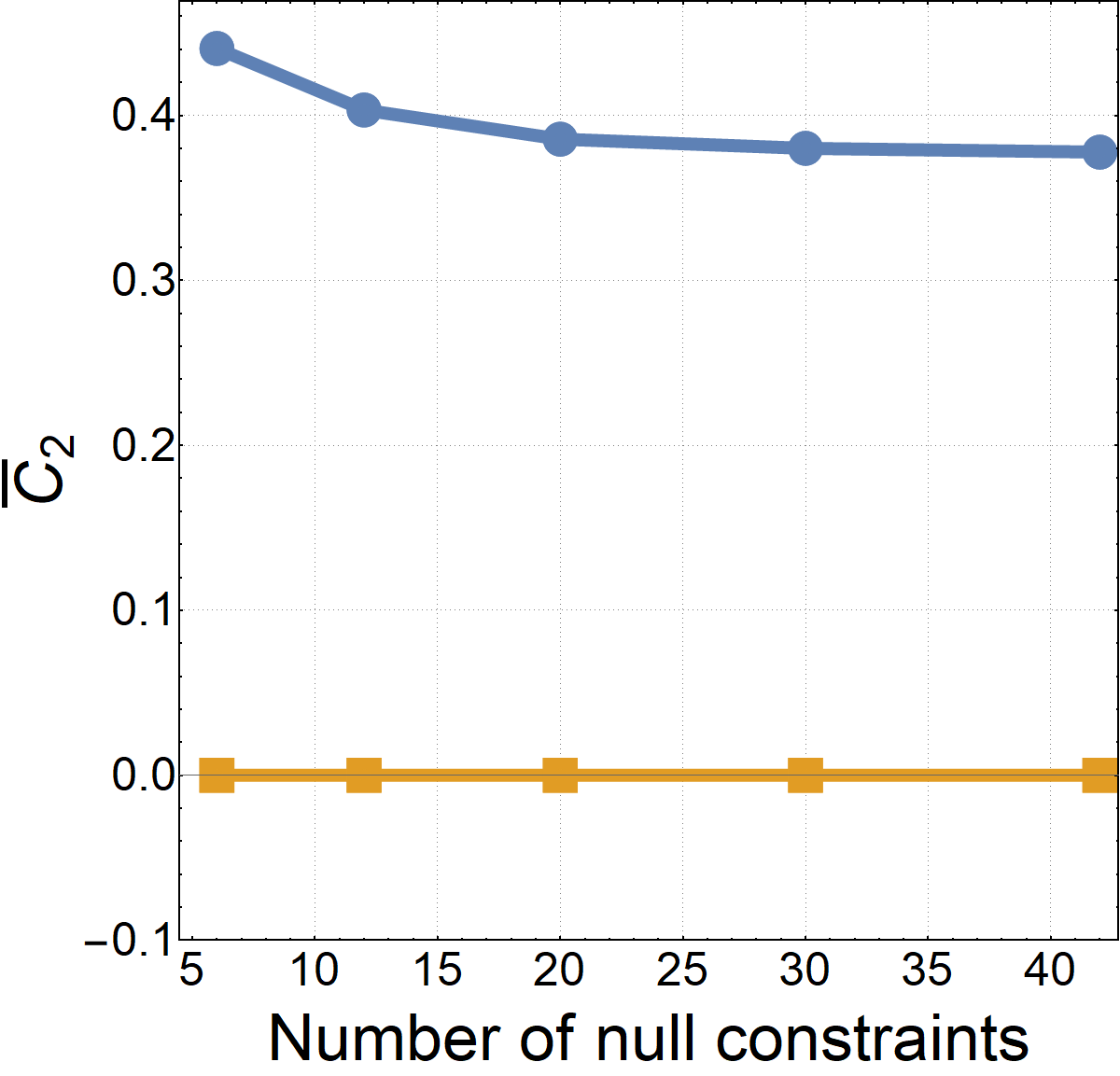}
    \end{subfigure}
    \begin{subfigure}[b]{0.3\textwidth}
        \includegraphics[width=\textwidth]{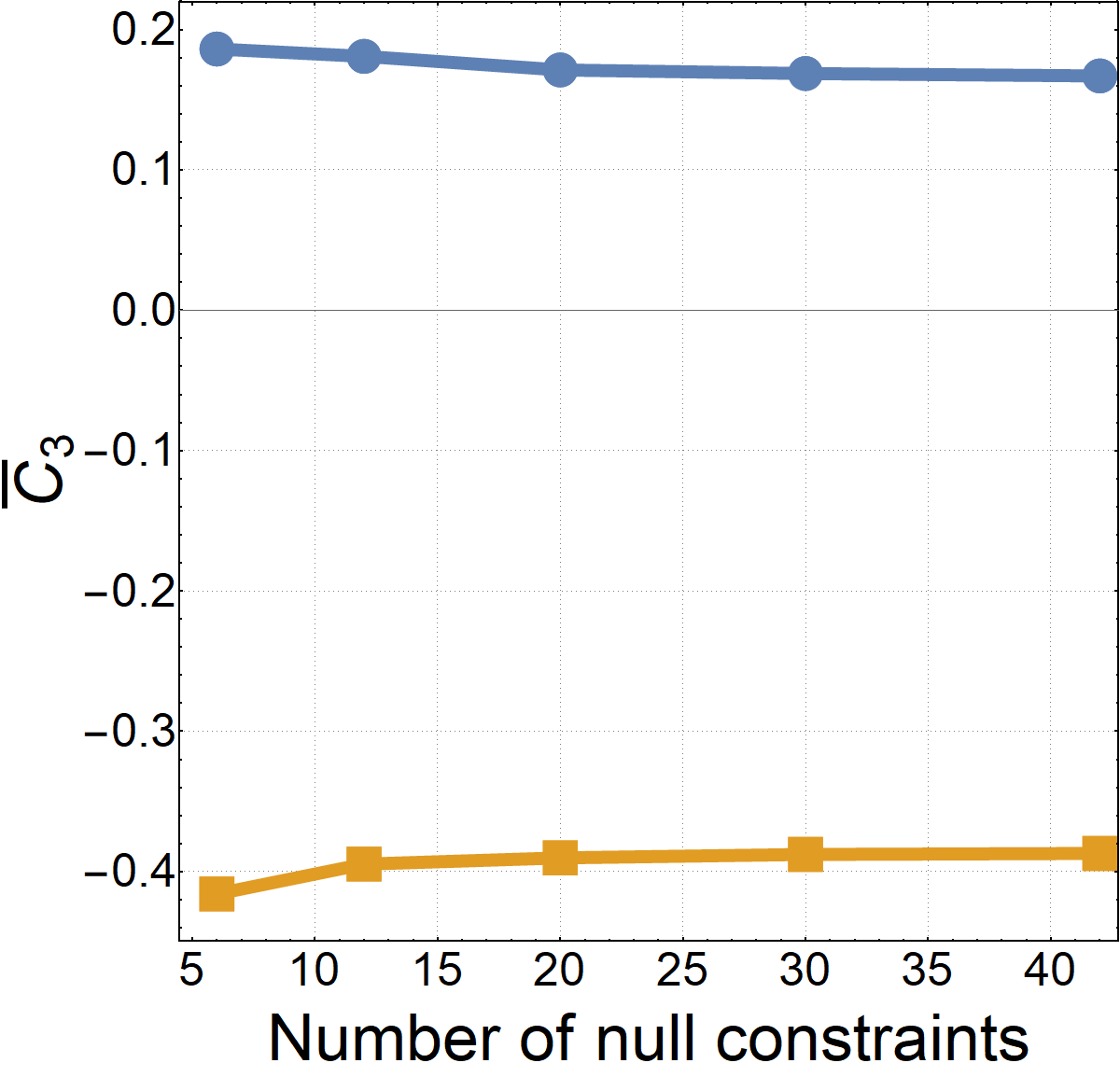}
    \end{subfigure}
    \caption{Convergence of positivity (upper and lower) bounds with the number of null constraint. We choose $N=10,\ell_M=20$.}
    \label{fig:null}
\end{figure}

\begin{figure}[tbp]
    \begin{subfigure}[b]{0.3\textwidth}
        \includegraphics[width=\textwidth]{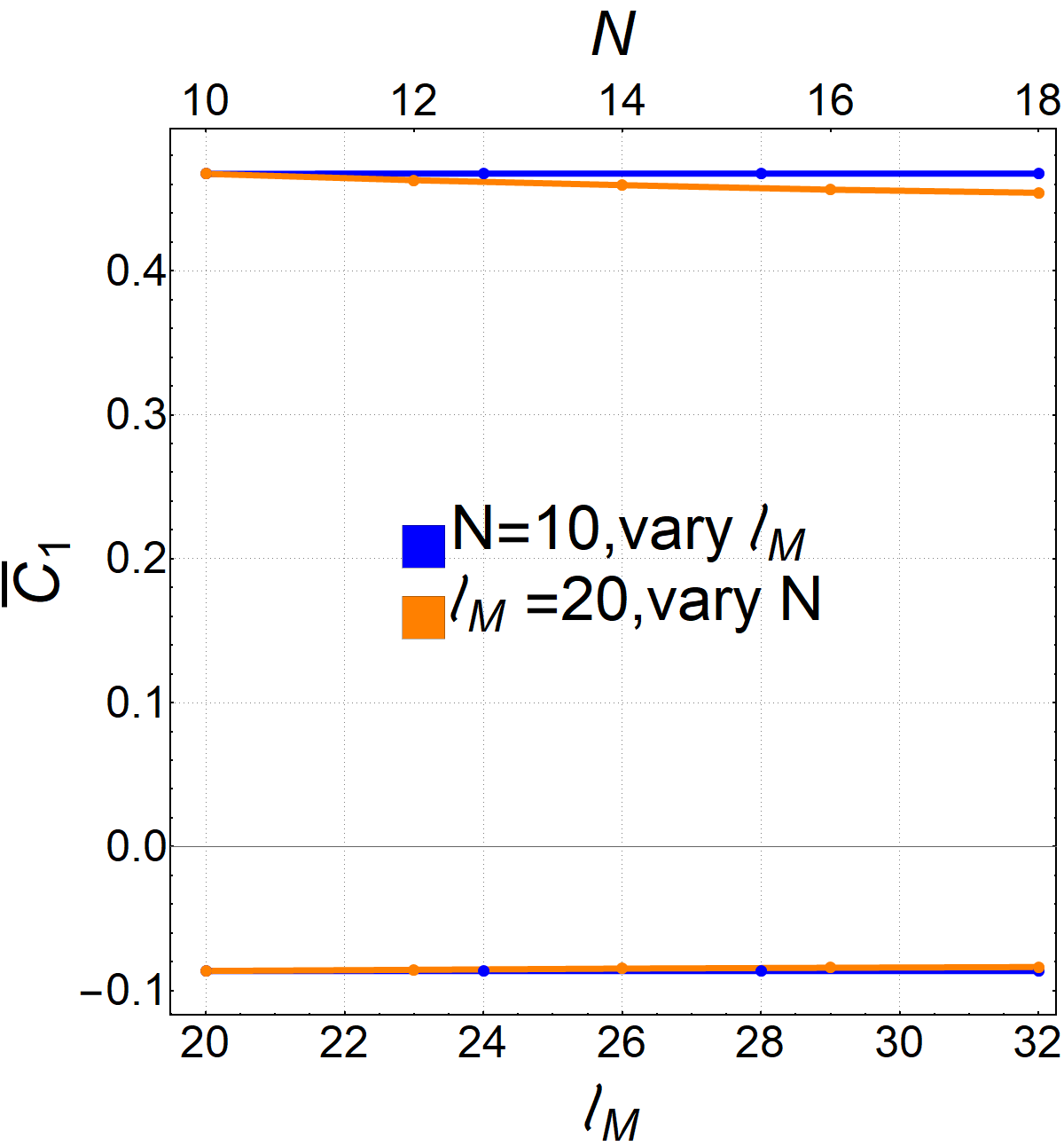}
    \end{subfigure}
    \begin{subfigure}[b]{0.3\textwidth}
        \includegraphics[width=\textwidth]{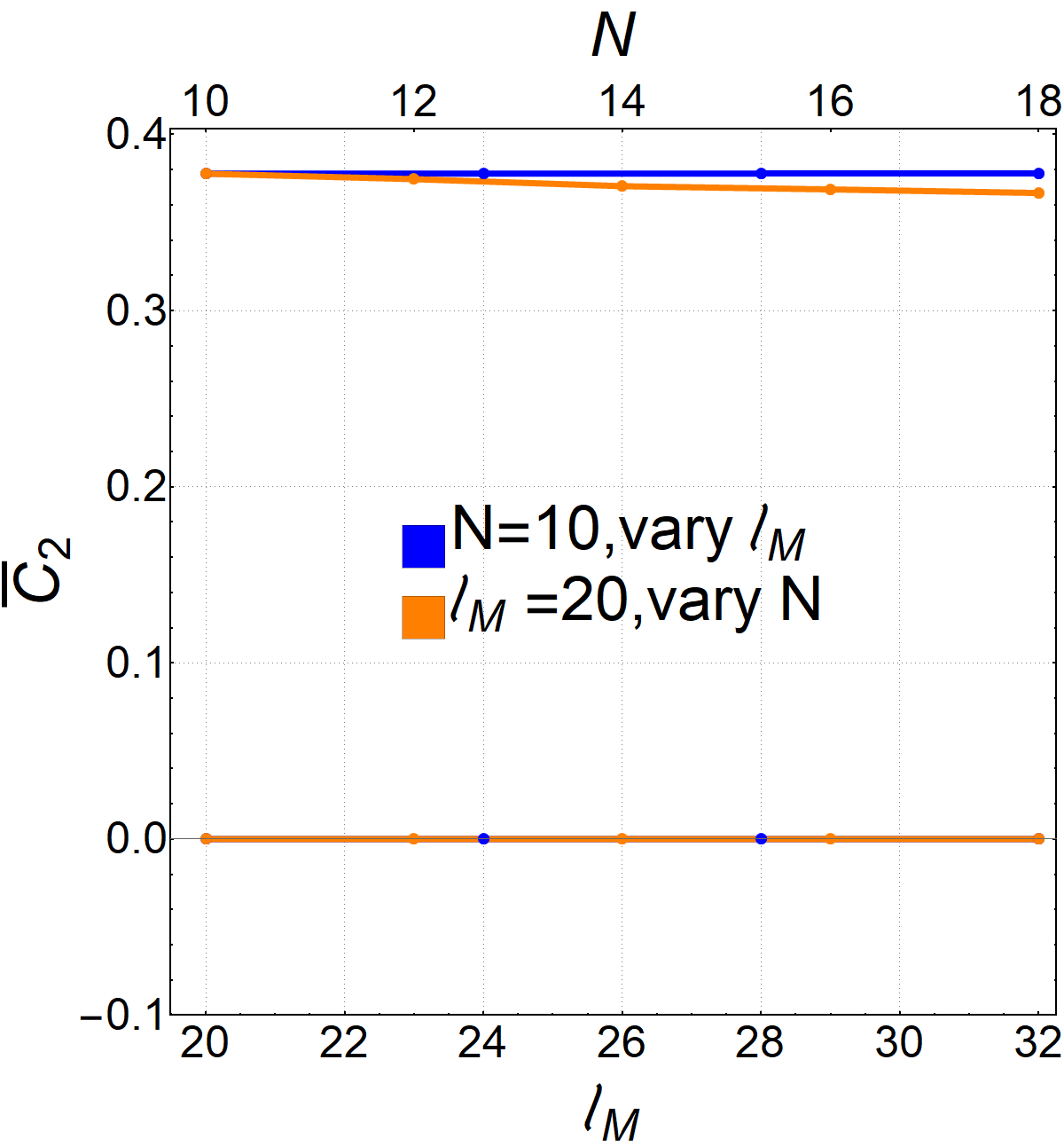}
    \end{subfigure}
    \begin{subfigure}[b]{0.3\textwidth}
        \includegraphics[width=\textwidth]{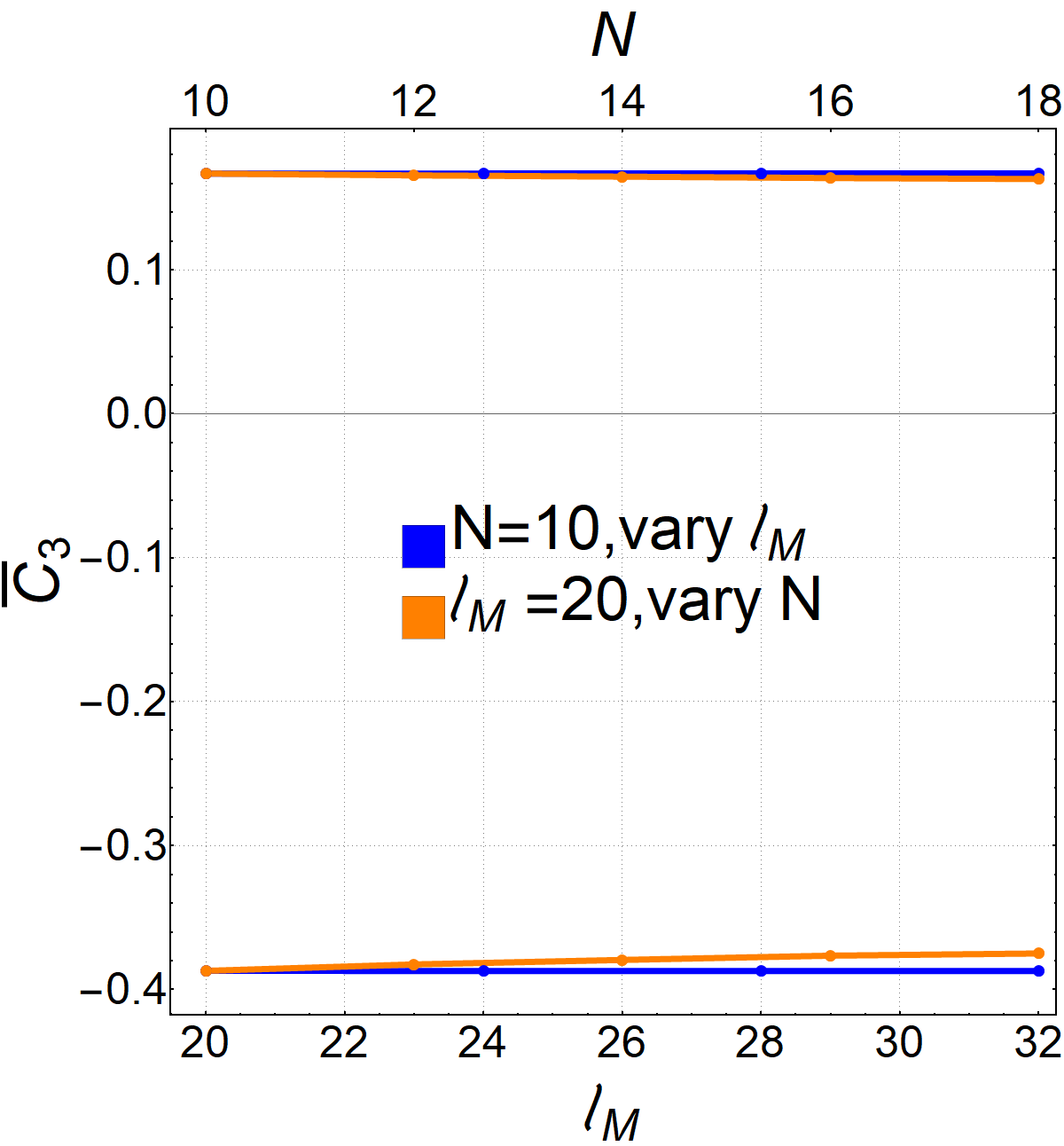}
    \end{subfigure}
    \caption{Convergence of positivity (upper and lower) bounds with the numerical truncations $\ell_M$ and $N$. Here $\bar{C_i}=C_i \Li^4 /(4\pi)^2$. 42 null constraints are used.}
    \label{fig:l}
\end{figure}

Finally, we would like to point out that the convergences of our numerically results are excellent. To see this, in Figure \ref{fig:null}, we plot how the 1D bounds varies with the number of null constraints used. In the above numerical results, we truncated the UV scales with $N=10$ and the UV spins with $\ell_M=20$, and we find that it is convenient to choose $\ell_{\infty}=100$. With this numerical setup, the computation of a single half-space bound uses about 110 CPU hours. As $N$ increases, the positivity bounds become weaker, while the bounds becomes tighter as $\ell_M$ increases. In Figure \ref{fig:l}, we see that the results are quite stable against increasing the values of $N$ and $\ell_M$.

\section{Summary}

Positivity bounds are a set of highly restrictive conditions on the low-energy Wilson coefficients that have yet to be fully appreciated by the wider particle phenomenological and experimental communities. Although the formalism of positivity bounds itself is still under active development, highly constraining results are already available and straightforward to use. Here we take the Higgs scattering in the SMEFT as an example to illustrate how to numerically compute optimal, two-sided positivity bounds on ths dimension-8 Wilson coefficients. While the SMEFT formalism is generic, we assume that the SMEFT is weakly coupled below the EFT cutoff but may be strongly coupled in the UV. The formalism presented can be easily generalized to other sectors of the SMEFT, which is left for future work.

In this paper, we have improved the existing positivity bounds on the SMEFT Higgs by applying nonlinear unitarity conditions to the UV amplitude or spectral functions and by leveraging the full internal symmetries of the Higgs scattering. While the previous bounds can be obtained by simple {\tt Mathematica} coding with linear programming, our new results make use of the {\tt SDPB} package, which can solve various field-theoretical semi-definite programs efficiently and highly accurately. We see that the new bounds are significantly stronger.

We have found that, in the Higgs case, the robust internal symmetries imply that the linear UV unitarity conditions used in Ref \cite{Chen:2023bhu} are actually tantamount to the nonlinear unitarity conditions. However, including the full internal symmetries does lead to tighter positivity bounds than the previous ones. As these new bounds are stronger than the current experimental bounds and the partial wave unitarity bounds, they will be useful in analyzing the current and upcoming phenomenological data for dimension-8 operators. These bounds may also be used to test the fundamental principles of quantum field theory or rule out UV particles from the collider data along the lines of \cite{Fuks:2020ujk, Li:2022rag}.

In general, the nonlinear unitarity conditions are of course more stringent. To demonstrate that the nonlinear unitarity conditions generally give rise to stronger bounds, we have calculated the two-sided positivity bounds for $\mathbb{Z}_2$ bi-scalar theory, a theory with two real scalar fields endowed with the reflection symmetry $\varphi_i \rightarrow -\varphi_i, ~i=1,2$. Nevertheless, we see that the linear unitarity conditions already give rise to bounds that are close to the bounds from the nonlinear conditions.

\acknowledgments

We would like to thank Yue-Zhou Li, Shi-Lin Wan, Tong Wu and Guo-Dong Zhang for helpful discussions. SYZ acknowledges support from the Fundamental Research Funds for the Central Universities under grant No.~WK2030000036 and from the National Natural Science Foundation of China under grant No.~12075233.

\appendix

\section{Linear unitarity from nonlinear unitarity}
\label{sec:linear}

In this Appendix, we re-derive the linear unitarity conditions of Ref \cite{Chen:2023bhu} from the linear matrix inequalities (\ref{nonlinearUell}), which are nonlinear in terms of the partial wave amplitudes. In this appendix only, we choose the physical momenta for all the particles, not using the all in-going convention. 

Firstly, we derive the linear unitarity conditions on $\ri_\ell^{iiii}$ and $\ri_\ell^{iijj}$ with $i\neq j$. Let us focus on the following sub-matrix, which is the smallest sub-matrix containing $\ri_\ell^{iiii}$ and $\ri_\ell^{iijj}$.
\be 
\begin{pmatrix}
\ri_\ell^{iiii}  &  \ri_\ell^{iijj}  \\
\ri_\ell^{iijj}  &  \ri_\ell^{jjjj}
\end{pmatrix} .
\ee
If a Hermitian matrix M is positive semi-definite, then its principal sub-matrices are also positive semi-definite. So $\text{Im}\mathbb{T}\succeq 0$ and $2 \mathbb{I} - \text{Im}\mathbb{T} \succeq 0$ implies that we must have
\bal
\begin{pmatrix}
\ri_\ell^{iiii}  &  \ri_\ell^{iijj}  \\
\ri_\ell^{iijj}  &  \ri_\ell^{jjjj}
\end{pmatrix}
\succeq 0  ,~~~~
\begin{pmatrix}
2-\ri_\ell^{iiii}  &  -\ri_\ell^{iijj}  \\
-\ri_\ell^{iijj}  &  2-\ri_\ell^{jjjj}
\end{pmatrix}
\succeq 0    .
\eal
It is easy to see that these matrix inequalities imply
\be 
0\leq \ri_\ell^{iiii}  \leq 2 ,~~~~
\ri_\ell^{iiii} \ri_\ell^{jjjj} \geq \(\ri_\ell^{iijj}\)^2  ,~~~
\(2-\ri_\ell^{iiii}\)\(2- \ri_\ell^{jjjj}\) \geq \(\ri_\ell^{iijj}\)^2 .
\ee
Combining these with the arithmetic-geometric mean inequality, we immediately get
\bal
\f {\ri_\ell^{iiii}+\ri_\ell^{jjjj}} 2  \geq |\ri_\ell^{iijj}|,~~~~
\f {4-\ri_\ell^{iiii}-\ri_\ell^{jjjj}} 2  \geq |\ri_\ell^{iijj}| .
\eal

Next, we will derive the linear unitarity conditions on $\ri_\ell^{ijij},\ri_\ell^{klkl}$ and $\ri_\ell^{ijkl}$, where $i\ne j\ne k\ne l$. This time we consider the following sub-matrix of ${\rm Im}\mathbb{T}$, which only contains the two-particle states $ij,ji,kl,lk$,
\be 
\begin{pmatrix}
2\ri_\ell^{ijij}  & 2\ri_\ell^{ijji}  & 2\ri_\ell^{ijkl} & 2\ri_\ell^{ijlk}   \\
2\ri_\ell^{jiij}  & 2\ri_\ell^{jiji}  & 2\ri_\ell^{jikl} & 2\ri_\ell^{jilk}   \\
2\ri_\ell^{klij}  & 2\ri_\ell^{klji}  & 2\ri_\ell^{klkl} & 2\ri_\ell^{kllk}   \\
2\ri_\ell^{lkij}  & 2\ri_\ell^{lkji}  & 2\ri_\ell^{lkkl} & 2\ri_\ell^{lklk}   
\end{pmatrix}
=
\begin{pmatrix}
2\ri_\ell^{ijij}  & 2\ri_\ell^{ijji}  & 2\ri_\ell^{ijkl} & 2\ri_\ell^{ijlk}   \\
2\ri_\ell^{ijji}  & 2\ri_\ell^{ijij}  & 2\ri_\ell^{ijlk} & 2\ri_\ell^{ijkl}   \\
2\ri_\ell^{ijkl}  & 2\ri_\ell^{ijlk}  & 2\ri_\ell^{klkl} & 2\ri_\ell^{kllk}   \\
2\ri_\ell^{ijlk}  & 2\ri_\ell^{ijkl}  & 2\ri_\ell^{kllk} & 2\ri_\ell^{klkl}   
\end{pmatrix} .
\label{bigmatrix1}
\ee
where the right hand side of the equality is obtained by using the relation $a_\ell^{i j k l}=(-1)^\ell a_\ell^{i j l k}$. If we further restrict to the upper-left $2\times2$ sub-matrix, which is a principal minor of ${\rm Im}\mathbb{T}$, \eqref{nonlinearUell} implies that
\bal
0\leq \ri_\ell^{ijij} \leq \f 1 2 ,
\eal
Similarly, if we consider the central $2\times2$ sub-matrix, ${\rm Im}\mathbb{T}\succeq 0$ leads to
\bal
\f {\ri_\ell^{ijij}+\ri_\ell^{klkl}} 2 \geq \sqrt{\ri_\ell^{ijij} \ri_\ell^{klkl}} \geq |\ri_\ell^{ijkl}|  . 
\eal
Furthermore, the second equation in \eqref{nonlinearUell} implies that the determinant of the full sub-matrix \eqref{bigmatrix1} is positive semi-definite
\bal
0&\leq
\text{Det}
\begin{pmatrix}
2-2\ri_\ell^{ijij}  & -2\ri_\ell^{ijji}  & -2\ri_\ell^{ijkl} & -2\ri_\ell^{ijlk}   \\
-2\ri_\ell^{ijji}  & 2-2\ri_\ell^{ijij}  & -2\ri_\ell^{ijlk} & -2\ri_\ell^{ijkl}   \\
-2\ri_\ell^{ijkl}  & -2\ri_\ell^{ijlk}  & 2-2\ri_\ell^{klkl} & -2\ri_\ell^{kllk}   \\
-2\ri_\ell^{ijlk}  & -2\ri_\ell^{ijkl}  & -2\ri_\ell^{kllk} & 2-2\ri_\ell^{klkl}   
\end{pmatrix}
\nn
&=16 - 32\ri_\ell^{ijij} - 32\ri_\ell^{klkl}  + 64\ri_\ell^{ijij}\ri_\ell^{klkl}  - 64\(\ri_\ell^{ijkl}\)^2    \nn
&\leq 16 - 32\ri_\ell^{ijij} - 32\ri_\ell^{klkl}  + 16\(\ri_\ell^{ijij}+\ri_\ell^{klkl}\)^2  - 64\(\ri_\ell^{ijkl}\)^2 \nn
&=16\(1-\ri_\ell^{ijij}-\ri_\ell^{klkl}\)^2-64\(\ri_\ell^{ijkl}\)^2  ,
\eal
which in turn leads to 
\be
 |\ri_\ell^{ijkl}|\leq \f{1-\ri_\ell^{ijij}-\ri_\ell^{klkl}} 2 .
\ee
For the linear inequality $|(\ri_\ell^{iijj}+\ri_\ell^{kkll})\pm (\ri_\ell^{iikk}+\ri_\ell^{jjll})|\leq 2$, it is not straightforward to see from the nonlinear conditions $\text{Im}\mathbb{T}\succeq 0$ and $2 \mathbb{I} - \text{Im}\mathbb{T} \succeq 0$. For the Higgs case, with all the internal symmetries included, this inequality actually does not lead to extra constraints. Thus, we see that the nonlinear unitarity conditions we use in this paper are stronger than the linear unitarity conditions used in Ref \cite{Chen:2023bhu}.

\section{Bounds on $\mathbb{Z}_2$ bi-scalar theory}
\label{sec:biS}

In this appendix, we take the $\mathbb{Z}_2$ bi-scalar theory as an example to illustrate that generically, in the absence of strong symmetries, the nonlinear unitarity conditions, as expected, do lead to stronger positivity bounds than the linear unitarity conditions of \cite{Chen:2023bhu}.

By $\mathbb{Z}_2$ bi-scalar theory, we mean a theory with two real scalar fields, $\varphi_1$ and $\varphi_2$, where the theory is invariant under the $\mathbb{Z}_2$ symmetry $\varphi_i \rightarrow -\varphi_i, ~i=1,2$. It is easy to see that in $\mathbb{Z}_2$ bi-scalar theory, 2-to-2 amplitudes $\mc{A}^{1112}$ and $\mc{A}^{2221}$ vanish, the same also applicable to the amplitudes with the cyclic permutations of $\mc{A}^{1112}$ and $\mc{A}^{2221}$. Taking these into account, we can write the amplitudes as follows
\bal
\mc{A}^{1111}(s,t)&=g^a_{0,0}+g^a_{1,0}\(s^2+t^2+u^2\)+g^a_{0,1}stu+\cdots   \\
\mc{A}^{1122}(s,t)&=g^b_{0,0}+g^b_{1,0}s+g^b_{0,1}tu+g^b_{2,0}s^2+g^b_{1,1}stu+g^b_{3,0}s^3+\cdots   \\
\mc{A}^{2222}(s,t)&=g^c_{0,0}+g^c_{1,0}\(s^2+t^2+u^2\)+g^c_{0,1}stu+\cdots 
\eal
and the nonlinear unitarity conditions are given by
\bal
\begin{pmatrix}
\ri_\ell^{1111}  &  \ri_\ell^{1122}  &  0  &  0 \\
\ri_\ell^{2211}  &  \ri_\ell^{2222}  &  0  &  0 \\
0  &  0  &  2\ri_\ell^{1212}  &  2\ri_\ell^{1221} \\
0  &  0  &  2\ri_\ell^{2112}  &  2\ri_\ell^{2121} \\
\end{pmatrix}
\succeq 0   ,~~~~
\begin{pmatrix}
2-\ri_\ell^{1111}  &  -\ri_\ell^{1122}  &  0  &  0 \\
-\ri_\ell^{2211}  &  2-\ri_\ell^{2222}  &  0  &  0 \\
0  &  0  &  2-2\ri_\ell^{1212}  &  -2\ri_\ell^{1221} \\
0  &  0  &  -2\ri_\ell^{2112}  &  2-2\ri_\ell^{2121} \\
\end{pmatrix}
\succeq 0 ,
\eal
which is equivalent to
\bal
\label{biun3}
0\leq \ri_\ell^{1111}&\leq 2 ,&~~~~
0\leq \ri_\ell^{2222}&\leq 2 ,&~~~~
0\leq \ri_\ell^{1212}&\leq \f 1 2,\\
\label{biun1}
\ri_\ell^{1111}\ri_\ell^{2222}&\ge\(\ri_\ell^{1122}\)^2  ,&~~~~
\(2-\ri_\ell^{1111}\)\(2-\ri_\ell^{2222}\)&\ge\(\ri_\ell^{1122}\)^2 .
\eal
The null constraints are just the same as \cite{Du:2021byy, Chen:2023bhu}. For our purposes, we will only calculate the positivity bounds on the following two coefficients as well as their sum as an example:
\bal
g^a_{1,0}&= \left \langle\f{\ri^{1111}_\ell(\mu)}{\mu^3}\right \rangle ,    \\
g^b_{2,0}&= \left \langle\f{\ri^{1122}_\ell(\mu) + (-1)^\ell \ri^{1212}_\ell(\mu)}{\mu^3}\right \rangle  .
\eal

For the numerical optimization scheme, we essentially follow the same scheme as the Higgs case in the main text. With similar notations, the SDP we need to solve is given by:
\bal
&\text{\textbf{Decision variables}}   \nn
&\qquad \ri^{1111}_{\ell,n},~\ri^{2222}_{\ell,n},~\ri^{1122}_{\ell,n},~\ri^{1212}_{\ell,n} \text{\qquad for $\ell=0,1,\dots,\ell_{\text{M}},\ell_{\infty}$ and $n= 1,2,..., N$}   \\
&\text{\textbf{Maximize/Minimize}}   \nn
&\qquad g^a_{1,0}=\f 1 {\Li^4}\sum_{\ell=0}^{\ell_{\text{M}},\ell_{\infty}} 16(2 \ell+1)\sum_{n=1}^{N}\f 1 N \f n N \ri^{1111}_{\ell,n}  \\
&\qquad g^b_{2,0}=\f 1 {\Li^4}\sum_{\ell=0}^{\ell_{\text{M}},\ell_{\infty}} 16(2 \ell+1)\sum_{n=1}^{N}\f 1 N \f n N \(\ri^{1122}_{\ell,n}+(-1)^\ell \ri^{1212}_{\ell,n}\)  \\
&\qquad g^a_{1,0}+g^b_{2,0}=\f 1 {\Li^4}\sum_{\ell=0}^{\ell_{\text{M}},\ell_{\infty}} 16(2 \ell+1)\sum_{n=1}^{N}\f 1 N \f n N \(\ri^{1111}_{\ell,n}+\ri^{1122}_{\ell,n}+(-1)^\ell \ri^{1212}_{\ell,n}\)  \\
&\text{\textbf{Subject to}}    \nn
&\qquad \text{Unitarity conditions }\eqref{biun3}\text{-}\eqref{biun1} \nn
&\qquad \text{Null constraints}  \nonumber
\eal

As for the corresponding linear unitarity conditions, it is easy to get them from \eqref{biun1} using the arithmetic-geometric mean inequality:
\bal
\label{biun5}
\f{\ri_\ell^{1111}+\ri_\ell^{2222}} 2 \ge |\ri_\ell^{1122}| \,,~~~~
\f{4-\ri_\ell^{1111}-\ri_\ell^{2222}} 2 \ge |\ri_\ell^{1122}|   .
\eal
To obtain the positivity bounds using linear unitarity conditions, we only need to replace the nonlinear unitarity conditions \eqref{biun1} by \eqref{biun5} in the above SDP. 

Note that the equality of the arithmetic-geometric mean inequality can be saturated when $\ri_\ell^{1111}=\ri_\ell^{2222}$. Thus, if we have  $\ri_\ell^{1111}=\ri_\ell^{2222}$ at (the boundaries of) the positivity bounds, the bounds from the linear unitarity conditions will be the same as those from the nonlinear unitarity conditions. It is for this reason that for a double $\mathbb{Z}_2$ bi-scalar theory where there is an additional $\mathbb{Z}_2$ symmetry $\varphi_1 \leftrightarrow \varphi_2$, the positivity bounds from the linear unitarity conditions are the same as those from the nonlinear unitarity conditions, which we have verified numerically.

\begin{table}
    \centering
    \begin{tabular}{|c|c|c|c|c|c|c|}
    \hline
                &  \multicolumn{2}{|c|}{$\bar{g}^a_{1,0}=g^a_{1,0} \Li^4 /(4\pi)^2$}  &  \multicolumn{2}{|c|}{$\bar{g}^b_{2,0}=g^b_{2,0} \Li^4 /(4\pi)^2$}  &  \multicolumn{2}{|c|}{$\bar{g}^a_{1,0}+\bar{g}^b_{2,0}$} \\
    \cline{2-7}
                & lower &  upper & lower & upper& lower & upper\\
    \hline
    Linear      & $0$ &$0.798$  & $-0.330$  & $0.614$ & $-0.326$ & $1.41$\\
    \hline
    Nonlinear   & $0$ &$0.798$   &$-0.330$  &$0.614$  &$-0.172$  &$1.18$ \\
    \hline
    \end{tabular}
    \caption{Bounds on $g^a_{1,0},g^b_{2,0}$ and $g^a_{1,0}+g^b_{2,0}$, using linear and nonlinear unitarity conditions separately. Here we choose $N=20,\ell_M=30$.}
    \label{tab:biscalar}
\end{table}

\begin{figure}[tbp]
\centering
    \includegraphics[width=0.5\textwidth]{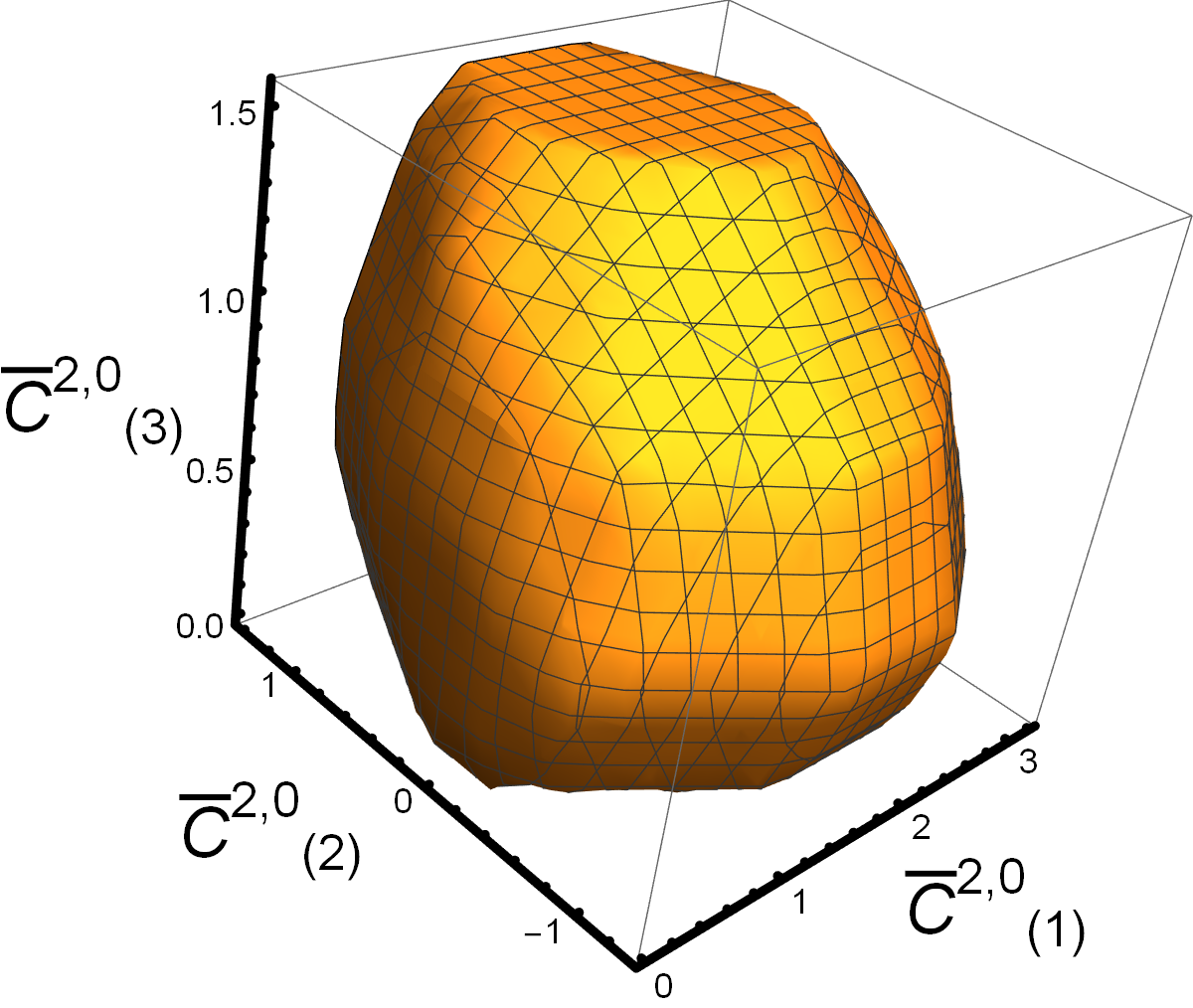}
    \caption{Positivity regions in the 3D subspaces of $c^{2,0}_{(1)}, c^{2,0}_{(2)}$ and $c^{2,0}_{(3)}$. Here, $\bar{c}^{2,0}_{(i)}=c^{2,0}_{(i)}\Li^4/(4\pi)^2$ and we choose $N=20,\ell_M=30$.}
    \label{fig:3d}
\end{figure}

We present the bounds on $g^a_{1,0},g^b_{2,0}$ and $g^a_{1,0}+g^b_{2,0}$ in Table \ref{tab:biscalar}. We observe that the linear and nonlinear bounds on $g^a_{1,0}+g^b_{2,0}$ are different, while those on $g^a_{1,0}$ and $g^b_{2,0}$ are the same. Note that the partial wave amplitude expansion of $g^a_{1,0}+g^b_{2,0}$ is not symmetric under the transformation $\ri_\ell^{1111}\leftrightarrow \ri_\ell^{2222}$. Thus, we can predict that the extrema of the $g^a_{1,0}+g^b_{2,0}$ bounds are achieved when $\ri_\ell^{1111}\ne \ri_\ell^{2222}$, leading to different bounds using the linear and nonlinear unitarity conditions. The $g^b_{2,0}$ case is just the opposite. As for $g^a_{1,0}$, its extrema can be achieved by setting $\ri^{1122}_\ell=0$, in which case \eqref{biun1} is automatically satisfied due to \eqref{biun3}. Thus, for $g^a_{1,0}$, the linear and nonlinear unitarity conditions are the same, which leads to the same positivity bounds.

For completeness, we also calculate the bounds on the following coefficients in a double $\mathbb{Z}_2$ symmetry bi-scalar theory:
\bal
%c^{2,0}_{1111} &\equiv 2g^a_{1,0} = \left \langle\f{2\ri^{1111}_\ell(\mu)}{\mu^3}\right \rangle ,
c^{2,0}_{(1)} &\equiv 4g^a_{1,0} = \left \langle\f{4\ri^{1111}_\ell(\mu)}{\mu^3}\right \rangle ,
\\
%c^{2,0}_{1122} &\equiv g^b_{2,0}= \left \langle\f{\ri^{1122}_\ell(\mu) + (-1)^\ell \ri^{1212}_\ell(\mu)}{\mu^3}\right \rangle  ,
c^{2,0}_{(2)} &\equiv 4g^b_{2,0}= \left \langle\f{4\ri^{1122}_\ell(\mu) + 4(-1)^\ell \ri^{1212}_\ell(\mu)}{\mu^3}\right \rangle  ,
\\
%c^{2,0}_{1212} &\equiv 2g^b_{2,0}+g^b_{0,1} = \left \langle\f{2\ri^{1212}_\ell(\mu)}{\mu^3}\right \rangle .
c^{2,0}_{(3)} &\equiv 4g^b_{2,0}+2g^b_{0,1} = \left \langle\f{4\ri^{1212}_\ell(\mu)}{\mu^3}\right \rangle .
\eal
The 3D bounds on these coefficients are shown in Figure \ref{fig:3d}, where we defined $\bar{c}^{2,0}_{(i)}=c^{2,0}_{(i)}\Li^4/(4\pi)^2$ for $i=1,2,3$. The two-sided bounds on these 3 coefficients
%and the bounds on a couple of specific directions 
are given by
\bal
0\leq \bar{c}^{2,0}_{(1)} \leq 3.19 \,, \quad 
-1.32 \leq \bar{c}^{2,0}_{(2)} \leq 2.46\,, \quad
0 \leq \bar{c}^{2,0}_{(3)} \leq 1.54   \,.
%\\
%\bar{c}^{2,0}_{(1)} \leq 3.192 \,, \quad \bar{c}^{2,0}_{(1)}+\bar{c}^{2,0}_{(2)}+\bar{c}^{2,0}_{(3)} \leq 4.942 \,, \quad \bar{c}^{2,0}_{(1)}+\bar{c}^{2,0}_{(3)} \leq  4.325  \,.
\eal
Of course, these numerical results are obtained with a relatively crude discretization scheme and only a few null constraints, and can be further improved.

\bibliographystyle{JHEP}
\bibliography{refs}

\providecommand{\href}[2]{#2}\begingroup\raggedright\begin{thebibliography}{10}

\bibitem{Adams:2006sv}
A.~Adams, N.~Arkani-Hamed, S.~Dubovsky, A.~Nicolis and R.~Rattazzi,
  \emph{{Causality, analyticity and an IR obstruction to UV completion}},
  \href{http://dx.doi.org/10.1088/1126-6708/2006/10/014}{\emph{JHEP} {\bfseries
  10} (2006) 014}, [\href{https://arxiv.org/abs/hep-th/0602178}{{\ttfamily
  hep-th/0602178}}].

\bibitem{deRham:2017avq}
C.~de~Rham, S.~Melville, A.~J. Tolley and S.-Y. Zhou, \emph{{Positivity bounds
  for scalar field theories}},
  \href{http://dx.doi.org/10.1103/PhysRevD.96.081702}{\emph{Phys. Rev. D}
  {\bfseries 96} (2017) 081702},
  [\href{https://arxiv.org/abs/1702.06134}{{\ttfamily 1702.06134}}].

\bibitem{deRham:2017zjm}
C.~de~Rham, S.~Melville, A.~J. Tolley and S.-Y. Zhou, \emph{{UV complete me:
  Positivity Bounds for Particles with Spin}},
  \href{http://dx.doi.org/10.1007/JHEP03(2018)011}{\emph{JHEP} {\bfseries 03}
  (2018) 011}, [\href{https://arxiv.org/abs/1706.02712}{{\ttfamily
  1706.02712}}].

\bibitem{Arkani-Hamed:2020blm}
N.~Arkani-Hamed, T.-C. Huang and Y.-t. Huang, \emph{{The EFT-Hedron}},
  \href{http://dx.doi.org/10.1007/JHEP05(2021)259}{\emph{JHEP} {\bfseries 05}
  (2021) 259}, [\href{https://arxiv.org/abs/2012.15849}{{\ttfamily
  2012.15849}}].

\bibitem{Bellazzini:2020cot}
B.~Bellazzini, J.~Elias~Mir\'o, R.~Rattazzi, M.~Riembau and F.~Riva,
  \emph{{Positive moments for scattering amplitudes}},
  \href{http://dx.doi.org/10.1103/PhysRevD.104.036006}{\emph{Phys. Rev. D}
  {\bfseries 104} (2021) 036006},
  [\href{https://arxiv.org/abs/2011.00037}{{\ttfamily 2011.00037}}].

\bibitem{Tolley:2020gtv}
A.~J. Tolley, Z.-Y. Wang and S.-Y. Zhou, \emph{{New positivity bounds from full
  crossing symmetry}},
  \href{http://dx.doi.org/10.1007/JHEP05(2021)255}{\emph{JHEP} {\bfseries 05}
  (2021) 255}, [\href{https://arxiv.org/abs/2011.02400}{{\ttfamily
  2011.02400}}].

\bibitem{Caron-Huot:2020cmc}
S.~Caron-Huot and V.~Van~Duong, \emph{{Extremal Effective Field Theories}},
  \href{http://dx.doi.org/10.1007/JHEP05(2021)280}{\emph{JHEP} {\bfseries 05}
  (2021) 280}, [\href{https://arxiv.org/abs/2011.02957}{{\ttfamily
  2011.02957}}].

\bibitem{Chiang:2021ziz}
L.-Y. Chiang, Y.-t. Huang, W.~Li, L.~Rodina and H.-C. Weng, \emph{{Into the
  EFThedron and UV constraints from IR consistency}},
  \href{http://dx.doi.org/10.1007/JHEP03(2022)063}{\emph{JHEP} {\bfseries 03}
  (2022) 063}, [\href{https://arxiv.org/abs/2105.02862}{{\ttfamily
  2105.02862}}].

\bibitem{Sinha:2020win}
A.~Sinha and A.~Zahed, \emph{{Crossing Symmetric Dispersion Relations in
  Quantum Field Theories}},
  \href{http://dx.doi.org/10.1103/PhysRevLett.126.181601}{\emph{Phys. Rev.
  Lett.} {\bfseries 126} (2021) 181601},
  [\href{https://arxiv.org/abs/2012.04877}{{\ttfamily 2012.04877}}].

\bibitem{Zhang:2020jyn}
C.~Zhang and S.-Y. Zhou, \emph{{Convex Geometry Perspective on the (Standard
  Model) Effective Field Theory Space}},
  \href{http://dx.doi.org/10.1103/PhysRevLett.125.201601}{\emph{Phys. Rev.
  Lett.} {\bfseries 125} (2020) 201601},
  [\href{https://arxiv.org/abs/2005.03047}{{\ttfamily 2005.03047}}].

\bibitem{Li:2021lpe}
X.~Li, H.~Xu, C.~Yang, C.~Zhang and S.-Y. Zhou, \emph{{Positivity in Multifield
  Effective Field Theories}},
  \href{http://dx.doi.org/10.1103/PhysRevLett.127.121601}{\emph{Phys. Rev.
  Lett.} {\bfseries 127} (2021) 121601},
  [\href{https://arxiv.org/abs/2101.01191}{{\ttfamily 2101.01191}}].

\bibitem{Bellazzini:2014waa}
B.~Bellazzini, L.~Martucci and R.~Torre, \emph{{Symmetries, Sum Rules and
  Constraints on Effective Field Theories}},
  \href{http://dx.doi.org/10.1007/JHEP09(2014)100}{\emph{JHEP} {\bfseries 09}
  (2014) 100}, [\href{https://arxiv.org/abs/1405.2960}{{\ttfamily 1405.2960}}].

\bibitem{Bellazzini:2016xrt}
B.~Bellazzini, \emph{{Softness and amplitudes\textquoteright{} positivity for
  spinning particles}},
  \href{http://dx.doi.org/10.1007/JHEP02(2017)034}{\emph{JHEP} {\bfseries 02}
  (2017) 034}, [\href{https://arxiv.org/abs/1605.06111}{{\ttfamily
  1605.06111}}].

\bibitem{Bern:2021ppb}
Z.~Bern, D.~Kosmopoulos and A.~Zhiboedov, \emph{{Gravitational effective field
  theory islands, low-spin dominance, and the four-graviton amplitude}},
  \href{http://dx.doi.org/10.1088/1751-8121/ac0e51}{\emph{J. Phys. A}
  {\bfseries 54} (2021) 344002},
  [\href{https://arxiv.org/abs/2103.12728}{{\ttfamily 2103.12728}}].

\bibitem{Alberte:2020jsk}
L.~Alberte, C.~de~Rham, S.~Jaitly and A.~J. Tolley, \emph{{Positivity Bounds
  and the Massless Spin-2 Pole}},
  \href{http://dx.doi.org/10.1103/PhysRevD.102.125023}{\emph{Phys. Rev. D}
  {\bfseries 102} (2020) 125023},
  [\href{https://arxiv.org/abs/2007.12667}{{\ttfamily 2007.12667}}].

\bibitem{Tokuda:2020mlf}
J.~Tokuda, K.~Aoki and S.~Hirano, \emph{{Gravitational positivity bounds}},
  \href{http://dx.doi.org/10.1007/JHEP11(2020)054}{\emph{JHEP} {\bfseries 11}
  (2020) 054}, [\href{https://arxiv.org/abs/2007.15009}{{\ttfamily
  2007.15009}}].

\bibitem{Caron-Huot:2021rmr}
S.~Caron-Huot, D.~Mazac, L.~Rastelli and D.~Simmons-Duffin, \emph{{Sharp
  boundaries for the swampland}},
  \href{http://dx.doi.org/10.1007/JHEP07(2021)110}{\emph{JHEP} {\bfseries 07}
  (2021) 110}, [\href{https://arxiv.org/abs/2102.08951}{{\ttfamily
  2102.08951}}].

\bibitem{Grall:2021xxm}
T.~Grall and S.~Melville, \emph{{Positivity bounds without boosts: New
  constraints on low energy effective field theories from the UV}},
  \href{http://dx.doi.org/10.1103/PhysRevD.105.L121301}{\emph{Phys. Rev. D}
  {\bfseries 105} (2022) L121301},
  [\href{https://arxiv.org/abs/2102.05683}{{\ttfamily 2102.05683}}].

\bibitem{Du:2021byy}
Z.-Z. Du, C.~Zhang and S.-Y. Zhou, \emph{{Triple crossing positivity bounds for
  multi-field theories}},
  \href{http://dx.doi.org/10.1007/JHEP12(2021)115}{\emph{JHEP} {\bfseries 12}
  (2021) 115}, [\href{https://arxiv.org/abs/2111.01169}{{\ttfamily
  2111.01169}}].

\bibitem{Alberte:2021dnj}
L.~Alberte, C.~de~Rham, S.~Jaitly and A.~J. Tolley, \emph{{Reverse
  Bootstrapping: IR Lessons for UV Physics}},
  \href{http://dx.doi.org/10.1103/PhysRevLett.128.051602}{\emph{Phys. Rev.
  Lett.} {\bfseries 128} (2022) 051602},
  [\href{https://arxiv.org/abs/2111.09226}{{\ttfamily 2111.09226}}].

\bibitem{Bellazzini:2021oaj}
B.~Bellazzini, M.~Riembau and F.~Riva, \emph{{IR side of positivity bounds}},
  \href{http://dx.doi.org/10.1103/PhysRevD.106.105008}{\emph{Phys. Rev. D}
  {\bfseries 106} (2022) 105008},
  [\href{https://arxiv.org/abs/2112.12561}{{\ttfamily 2112.12561}}].

\bibitem{Chowdhury:2021ynh}
S.~D. Chowdhury, K.~Ghosh, P.~Haldar, P.~Raman and A.~Sinha, \emph{{Crossing
  Symmetric Spinning S-matrix Bootstrap: EFT bounds}},
  \href{http://dx.doi.org/10.21468/SciPostPhys.13.3.051}{\emph{SciPost Phys.}
  {\bfseries 13} (2022) 051},
  [\href{https://arxiv.org/abs/2112.11755}{{\ttfamily 2112.11755}}].

\bibitem{Chiang:2022ltp}
L.-Y. Chiang, Y.-t. Huang, L.~Rodina and H.-C. Weng, \emph{{De-projecting the
  EFThedron}},  \href{https://arxiv.org/abs/2204.07140}{{\ttfamily
  2204.07140}}.

\bibitem{Caron-Huot:2022ugt}
S.~Caron-Huot, Y.-Z. Li, J.~Parra-Martinez and D.~Simmons-Duffin,
  \emph{{Causality constraints on corrections to Einstein gravity}},
  \href{http://dx.doi.org/10.1007/JHEP05(2023)122}{\emph{JHEP} {\bfseries 05}
  (2023) 122}, [\href{https://arxiv.org/abs/2201.06602}{{\ttfamily
  2201.06602}}].

\bibitem{Caron-Huot:2022jli}
S.~Caron-Huot, Y.-Z. Li, J.~Parra-Martinez and D.~Simmons-Duffin,
  \emph{{Graviton partial waves and causality in higher dimensions}},
  \href{http://dx.doi.org/10.1103/PhysRevD.108.026007}{\emph{Phys. Rev. D}
  {\bfseries 108} (2023) 026007},
  [\href{https://arxiv.org/abs/2205.01495}{{\ttfamily 2205.01495}}].

\bibitem{Henriksson:2022oeu}
J.~Henriksson, B.~McPeak, F.~Russo and A.~Vichi, \emph{{Bounding violations of
  the weak gravity conjecture}},
  \href{http://dx.doi.org/10.1007/JHEP08(2022)184}{\emph{JHEP} {\bfseries 08}
  (2022) 184}, [\href{https://arxiv.org/abs/2203.08164}{{\ttfamily
  2203.08164}}].

\bibitem{Chiang:2022jep}
L.-Y. Chiang, Y.-t. Huang, W.~Li, L.~Rodina and H.-C. Weng,
  \emph{{(Non)-projective bounds on gravitational EFT}},
  \href{https://arxiv.org/abs/2201.07177}{{\ttfamily 2201.07177}}.

\bibitem{Albert:2022oes}
J.~Albert and L.~Rastelli, \emph{{Bootstrapping pions at large N}},
  \href{http://dx.doi.org/10.1007/JHEP08(2022)151}{\emph{JHEP} {\bfseries 08}
  (2022) 151}, [\href{https://arxiv.org/abs/2203.11950}{{\ttfamily
  2203.11950}}].

\bibitem{CarrilloGonzalez:2023cbf}
M.~Carrillo~Gonz\'alez, C.~de~Rham, S.~Jaitly, V.~Pozsgay and A.~Tokareva,
  \emph{{Positivity-causality competition: a road to ultimate EFT consistency
  constraints}},  \href{https://arxiv.org/abs/2307.04784}{{\ttfamily
  2307.04784}}.

\bibitem{Hong:2023zgm}
D.-Y. Hong, Z.-H. Wang and S.-Y. Zhou, \emph{{Causality bounds on scalar-tensor
  EFTs}}, \href{http://dx.doi.org/10.1007/JHEP10(2023)135}{\emph{JHEP}
  {\bfseries 10} (2023) 135},
  [\href{https://arxiv.org/abs/2304.01259}{{\ttfamily 2304.01259}}].

\bibitem{Li:2023qzs}
Y.-Z. Li, \emph{{Effective field theory bootstrap, large-N $\chi$PT and
  holographic QCD}},  \href{https://arxiv.org/abs/2310.09698}{{\ttfamily
  2310.09698}}.

\bibitem{Paulos:2016but}
M.~F. Paulos, J.~Penedones, J.~Toledo, B.~C. van Rees and P.~Vieira, \emph{{The
  S-matrix bootstrap II: two dimensional amplitudes}},
  \href{http://dx.doi.org/10.1007/JHEP11(2017)143}{\emph{JHEP} {\bfseries 11}
  (2017) 143}, [\href{https://arxiv.org/abs/1607.06110}{{\ttfamily
  1607.06110}}].

\bibitem{Paulos:2017fhb}
M.~F. Paulos, J.~Penedones, J.~Toledo, B.~C. van Rees and P.~Vieira, \emph{{The
  S-matrix bootstrap. Part III: higher dimensional amplitudes}},
  \href{http://dx.doi.org/10.1007/JHEP12(2019)040}{\emph{JHEP} {\bfseries 12}
  (2019) 040}, [\href{https://arxiv.org/abs/1708.06765}{{\ttfamily
  1708.06765}}].

\bibitem{He:2018uxa}
Y.~He, A.~Irrgang and M.~Kruczenski, \emph{{A note on the S-matrix bootstrap
  for the 2d O(N) bosonic model}},
  \href{http://dx.doi.org/10.1007/JHEP11(2018)093}{\emph{JHEP} {\bfseries 11}
  (2018) 093}, [\href{https://arxiv.org/abs/1805.02812}{{\ttfamily
  1805.02812}}].

\bibitem{He:2021eqn}
Y.~He and M.~Kruczenski, \emph{{S-matrix bootstrap in 3+1 dimensions:
  regularization and dual convex problem}},
  \href{http://dx.doi.org/10.1007/JHEP08(2021)125}{\emph{JHEP} {\bfseries 08}
  (2021) 125}, [\href{https://arxiv.org/abs/2103.11484}{{\ttfamily
  2103.11484}}].

\bibitem{Karateev:2019ymz}
D.~Karateev, S.~Kuhn and J.~a. Penedones, \emph{{Bootstrapping Massive Quantum
  Field Theories}},
  \href{http://dx.doi.org/10.1007/JHEP07(2020)035}{\emph{JHEP} {\bfseries 07}
  (2020) 035}, [\href{https://arxiv.org/abs/1912.08940}{{\ttfamily
  1912.08940}}].

\bibitem{Guerrieri:2020bto}
A.~L. Guerrieri, J.~Penedones and P.~Vieira, \emph{{S-matrix bootstrap for
  effective field theories: massless pions}},
  \href{http://dx.doi.org/10.1007/JHEP06(2021)088}{\emph{JHEP} {\bfseries 06}
  (2021) 088}, [\href{https://arxiv.org/abs/2011.02802}{{\ttfamily
  2011.02802}}].

\bibitem{Kruczenski:2020ujw}
M.~Kruczenski and H.~Murali, \emph{{The R-matrix bootstrap for the 2d O(N)
  bosonic model with a boundary}},
  \href{http://dx.doi.org/10.1007/JHEP04(2021)097}{\emph{JHEP} {\bfseries 04}
  (2021) 097}, [\href{https://arxiv.org/abs/2012.15576}{{\ttfamily
  2012.15576}}].

\bibitem{Guerrieri:2021tak}
A.~Guerrieri and A.~Sever, \emph{{Rigorous Bounds on the Analytic S Matrix}},
  \href{http://dx.doi.org/10.1103/PhysRevLett.127.251601}{\emph{Phys. Rev.
  Lett.} {\bfseries 127} (2021) 251601},
  [\href{https://arxiv.org/abs/2106.10257}{{\ttfamily 2106.10257}}].

\bibitem{Guerrieri:2021ivu}
A.~Guerrieri, J.~Penedones and P.~Vieira, \emph{{Where Is String Theory in the
  Space of Scattering Amplitudes?}},
  \href{http://dx.doi.org/10.1103/PhysRevLett.127.081601}{\emph{Phys. Rev.
  Lett.} {\bfseries 127} (2021) 081601},
  [\href{https://arxiv.org/abs/2102.02847}{{\ttfamily 2102.02847}}].

\bibitem{Albert:2023jtd}
J.~Albert and L.~Rastelli, \emph{{Bootstrapping Pions at Large $N$. Part II:
  Background Gauge Fields and the Chiral Anomaly}},
  \href{https://arxiv.org/abs/2307.01246}{{\ttfamily 2307.01246}}.

\bibitem{Acanfora:2023axz}
F.~Acanfora, A.~Guerrieri, K.~H\"aring and D.~Karateev, \emph{{Bounds on
  scattering of neutral Goldstones}},
  \href{https://arxiv.org/abs/2310.06027}{{\ttfamily 2310.06027}}.

\bibitem{Miro:2023bon}
J.~Elias~Miro, A.~Guerrieri and M.~A. Gumus, \emph{{Extremal Higgs couplings}},
   \href{https://arxiv.org/abs/2311.09283}{{\ttfamily 2311.09283}}.

\bibitem{deRham:2022hpx}
C.~de~Rham, S.~Kundu, M.~Reece, A.~J. Tolley and S.-Y. Zhou, \emph{{Snowmass
  White Paper: UV Constraints on IR Physics}},  in \emph{{Snowmass 2021}}, 3,
  2022.
\newblock \href{https://arxiv.org/abs/2203.06805}{{\ttfamily 2203.06805}}.

\bibitem{Zhang:2018shp}
C.~Zhang and S.-Y. Zhou, \emph{{Positivity bounds on vector boson scattering at
  the LHC}}, \href{http://dx.doi.org/10.1103/PhysRevD.100.095003}{\emph{Phys.
  Rev. D} {\bfseries 100} (2019) 095003},
  [\href{https://arxiv.org/abs/1808.00010}{{\ttfamily 1808.00010}}].

\bibitem{Bi:2019phv}
Q.~Bi, C.~Zhang and S.-Y. Zhou, \emph{{Positivity constraints on aQGC: carving
  out the physical parameter space}},
  \href{http://dx.doi.org/10.1007/JHEP06(2019)137}{\emph{JHEP} {\bfseries 06}
  (2019) 137}, [\href{https://arxiv.org/abs/1902.08977}{{\ttfamily
  1902.08977}}].

\bibitem{Bellazzini:2018paj}
B.~Bellazzini and F.~Riva, \emph{{New phenomenological and theoretical
  perspective on anomalous ZZ and Z\ensuremath{\gamma} processes}},
  \href{http://dx.doi.org/10.1103/PhysRevD.98.095021}{\emph{Phys. Rev. D}
  {\bfseries 98} (2018) 095021},
  [\href{https://arxiv.org/abs/1806.09640}{{\ttfamily 1806.09640}}].

\bibitem{Remmen:2019cyz}
G.~N. Remmen and N.~L. Rodd, \emph{{Consistency of the Standard Model Effective
  Field Theory}}, \href{http://dx.doi.org/10.1007/JHEP12(2019)032}{\emph{JHEP}
  {\bfseries 12} (2019) 032},
  [\href{https://arxiv.org/abs/1908.09845}{{\ttfamily 1908.09845}}].

\bibitem{Yamashita:2020gtt}
K.~Yamashita, C.~Zhang and S.-Y. Zhou, \emph{{Elastic positivity vs extremal
  positivity bounds in SMEFT: a case study in transversal electroweak
  gauge-boson scatterings}},
  \href{http://dx.doi.org/10.1007/JHEP01(2021)095}{\emph{JHEP} {\bfseries 01}
  (2021) 095}, [\href{https://arxiv.org/abs/2009.04490}{{\ttfamily
  2009.04490}}].

\bibitem{Trott:2020ebl}
T.~Trott, \emph{{Causality, unitarity and symmetry in effective field theory}},
  \href{http://dx.doi.org/10.1007/JHEP07(2021)143}{\emph{JHEP} {\bfseries 07}
  (2021) 143}, [\href{https://arxiv.org/abs/2011.10058}{{\ttfamily
  2011.10058}}].

\bibitem{Remmen:2020vts}
G.~N. Remmen and N.~L. Rodd, \emph{{Flavor Constraints from Unitarity and
  Analyticity}},
  \href{http://dx.doi.org/10.1103/PhysRevLett.127.149901}{\emph{Phys. Rev.
  Lett.} {\bfseries 125} (2020) 081601},
  [\href{https://arxiv.org/abs/2004.02885}{{\ttfamily 2004.02885}}].

\bibitem{Remmen:2020uze}
G.~N. Remmen and N.~L. Rodd, \emph{{Signs, spin, SMEFT: Sum rules at dimension
  six}}, \href{http://dx.doi.org/10.1103/PhysRevD.105.036006}{\emph{Phys. Rev.
  D} {\bfseries 105} (2022) 036006},
  [\href{https://arxiv.org/abs/2010.04723}{{\ttfamily 2010.04723}}].

\bibitem{Gu:2020thj}
J.~Gu and L.-T. Wang, \emph{{Sum Rules in the Standard Model Effective Field
  Theory from Helicity Amplitudes}},
  \href{http://dx.doi.org/10.1007/JHEP03(2021)149}{\emph{JHEP} {\bfseries 03}
  (2021) 149}, [\href{https://arxiv.org/abs/2008.07551}{{\ttfamily
  2008.07551}}].

\bibitem{Fuks:2020ujk}
B.~Fuks, Y.~Liu, C.~Zhang and S.-Y. Zhou, \emph{{Positivity in
  electron-positron scattering: testing the axiomatic quantum field theory
  principles and probing the existence of UV states}},
  \href{http://dx.doi.org/10.1088/1674-1137/abcd8c}{\emph{Chin. Phys. C}
  {\bfseries 45} (2021) 023108},
  [\href{https://arxiv.org/abs/2009.02212}{{\ttfamily 2009.02212}}].

\bibitem{Gu:2020ldn}
J.~Gu, L.-T. Wang and C.~Zhang, \emph{{Unambiguously Testing Positivity at
  Lepton Colliders}},
  \href{http://dx.doi.org/10.1103/PhysRevLett.129.011805}{\emph{Phys. Rev.
  Lett.} {\bfseries 129} (2022) 011805},
  [\href{https://arxiv.org/abs/2011.03055}{{\ttfamily 2011.03055}}].

\bibitem{Bonnefoy:2020yee}
Q.~Bonnefoy, E.~Gendy and C.~Grojean, \emph{{Positivity bounds on Minimal
  Flavor Violation}},
  \href{http://dx.doi.org/10.1007/JHEP04(2021)115}{\emph{JHEP} {\bfseries 04}
  (2021) 115}, [\href{https://arxiv.org/abs/2011.12855}{{\ttfamily
  2011.12855}}].

\bibitem{Davighi:2021osh}
J.~Davighi, S.~Melville and T.~You, \emph{{Natural selection rules: new
  positivity bounds for massive spinning particles}},
  \href{http://dx.doi.org/10.1007/JHEP02(2022)167}{\emph{JHEP} {\bfseries 02}
  (2022) 167}, [\href{https://arxiv.org/abs/2108.06334}{{\ttfamily
  2108.06334}}].

\bibitem{Chala:2021wpj}
M.~Chala and J.~Santiago, \emph{{Positivity bounds in the standard model
  effective field theory beyond tree level}},
  \href{http://dx.doi.org/10.1103/PhysRevD.105.L111901}{\emph{Phys. Rev. D}
  {\bfseries 105} (2022) L111901},
  [\href{https://arxiv.org/abs/2110.01624}{{\ttfamily 2110.01624}}].

\bibitem{Zhang:2021eeo}
C.~Zhang, \emph{{SMEFTs living on the edge: determining the UV theories from
  positivity and extremality}},
  \href{http://dx.doi.org/10.1007/JHEP12(2022)096}{\emph{JHEP} {\bfseries 12}
  (2022) 096}, [\href{https://arxiv.org/abs/2112.11665}{{\ttfamily
  2112.11665}}].

\bibitem{Ghosh:2022qqq}
D.~Ghosh, R.~Sharma and F.~Ullah, \emph{{Amplitude\textquoteright{}s positivity
  vs. subluminality: causality and unitarity constraints on dimension 6 \& 8
  gluonic operators in the SMEFT}},
  \href{http://dx.doi.org/10.1007/JHEP02(2023)199}{\emph{JHEP} {\bfseries 02}
  (2023) 199}, [\href{https://arxiv.org/abs/2211.01322}{{\ttfamily
  2211.01322}}].

\bibitem{Remmen:2022orj}
G.~N. Remmen and N.~L. Rodd, \emph{{Spinning sum rules for the dimension-six
  SMEFT}}, \href{http://dx.doi.org/10.1007/JHEP09(2022)030}{\emph{JHEP}
  {\bfseries 09} (2022) 030},
  [\href{https://arxiv.org/abs/2206.13524}{{\ttfamily 2206.13524}}].

\bibitem{Li:2022tcz}
X.~Li and S.~Zhou, \emph{{Origin of neutrino masses on the convex cone of
  positivity bounds}},
  \href{http://dx.doi.org/10.1103/PhysRevD.107.L031902}{\emph{Phys. Rev. D}
  {\bfseries 107} (2023) L031902},
  [\href{https://arxiv.org/abs/2202.12907}{{\ttfamily 2202.12907}}].

\bibitem{Li:2022rag}
X.~Li, K.~Mimasu, K.~Yamashita, C.~Yang, C.~Zhang and S.-Y. Zhou,
  \emph{{Moments for positivity: using Drell-Yan data to test positivity bounds
  and reverse-engineer new physics}},
  \href{http://dx.doi.org/10.1007/JHEP10(2022)107}{\emph{JHEP} {\bfseries 10}
  (2022) 107}, [\href{https://arxiv.org/abs/2204.13121}{{\ttfamily
  2204.13121}}].

\bibitem{Li:2022aby}
X.~Li, \emph{{Positivity bounds at one-loop level: the Higgs sector}},
  \href{http://dx.doi.org/10.1007/JHEP05(2023)230}{\emph{JHEP} {\bfseries 05}
  (2023) 230}, [\href{https://arxiv.org/abs/2212.12227}{{\ttfamily
  2212.12227}}].

\bibitem{Altmannshofer:2023bfk}
W.~Altmannshofer, S.~Gori, B.~V. Lehmann and J.~Zuo, \emph{{UV physics from IR
  features: New prospects from top flavor violation}},
  \href{http://dx.doi.org/10.1103/PhysRevD.107.095025}{\emph{Phys. Rev. D}
  {\bfseries 107} (2023) 095025},
  [\href{https://arxiv.org/abs/2303.00781}{{\ttfamily 2303.00781}}].

\bibitem{Davighi:2023acq}
J.~Davighi, S.~Melville, K.~Mimasu and T.~You, \emph{{Positivity and the
  Electroweak Hierarchy}},  \href{https://arxiv.org/abs/2308.06226}{{\ttfamily
  2308.06226}}.

\bibitem{Ellis:2023zim}
J.~Ellis, K.~Mimasu and F.~Zampedri, \emph{{Dimension-8 SMEFT analysis of
  minimal scalar field extensions of the Standard Model}},
  \href{http://dx.doi.org/10.1007/JHEP10(2023)051}{\emph{JHEP} {\bfseries 10}
  (2023) 051}, [\href{https://arxiv.org/abs/2304.06663}{{\ttfamily
  2304.06663}}].

\bibitem{Chala:2023xjy}
M.~Chala and X.~Li, \emph{{Positivity restrictions on the mixing of
  dimension-eight SMEFT operators}},
  \href{https://arxiv.org/abs/2309.16611}{{\ttfamily 2309.16611}}.

\bibitem{Gu:2023emi}
J.~Gu and C.~Shu, \emph{{Probing positivity at the LHC with exclusive
  photon-fusion processes}},
  \href{https://arxiv.org/abs/2311.07663}{{\ttfamily 2311.07663}}.

\bibitem{Li:2020gnx}
H.-L. Li, Z.~Ren, J.~Shu, M.-L. Xiao, J.-H. Yu and Y.-H. Zheng, \emph{{Complete
  set of dimension-eight operators in the standard model effective field
  theory}}, \href{http://dx.doi.org/10.1103/PhysRevD.104.015026}{\emph{Phys.
  Rev. D} {\bfseries 104} (2021) 015026},
  [\href{https://arxiv.org/abs/2005.00008}{{\ttfamily 2005.00008}}].

\bibitem{Murphy:2020rsh}
C.~W. Murphy, \emph{{Dimension-8 operators in the Standard Model Eective Field
  Theory}}, \href{http://dx.doi.org/10.1007/JHEP10(2020)174}{\emph{JHEP}
  {\bfseries 10} (2020) 174},
  [\href{https://arxiv.org/abs/2005.00059}{{\ttfamily 2005.00059}}].

\bibitem{Vecchi:2007na}
L.~Vecchi, \emph{{Causal versus analytic constraints on anomalous quartic gauge
  couplings}},
  \href{http://dx.doi.org/10.1088/1126-6708/2007/11/054}{\emph{JHEP} {\bfseries
  11} (2007) 054}, [\href{https://arxiv.org/abs/0704.1900}{{\ttfamily
  0704.1900}}].

\bibitem{Chen:2023bhu}
Q.~Chen, K.~Mimasu, T.~A. Wu, G.-D. Zhang and S.-Y. Zhou, \emph{{Capping the
  positivity cone: dimension-8 Higgs operators in the SMEFT}},
  \href{https://arxiv.org/abs/2309.15922}{{\ttfamily 2309.15922}}.

\bibitem{Froissart:1961ux}
M.~Froissart, \emph{{Asymptotic behavior and subtractions in the Mandelstam
  representation}},
  \href{http://dx.doi.org/10.1103/PhysRev.123.1053}{\emph{Phys. Rev.}
  {\bfseries 123} (1961) 1053--1057}.

\bibitem{Martin:1962rt}
A.~Martin, \emph{{Unitarity and high-energy behavior of scattering
  amplitudes}}, \href{http://dx.doi.org/10.1103/PhysRev.129.1432}{\emph{Phys.
  Rev.} {\bfseries 129} (1963) 1432--1436}.

\bibitem{Landry:2019qug}
W.~Landry and D.~Simmons-Duffin, \emph{{Scaling the semidefinite program solver
  SDPB}},  \href{https://arxiv.org/abs/1909.09745}{{\ttfamily 1909.09745}}.

\end{thebibliography}\endgroup

\end{document}